\documentclass[prb,twocolumn,aps,superscriptaddress,floatfix]{revtex4}
\usepackage{amsmath}
\usepackage{amssymb}
\usepackage{bm}
\usepackage{graphicx}
\usepackage{color}
\usepackage[svgnames]{xcolor}
\usepackage{soul}
\usepackage[normalem]{ulem}
\usepackage{hyperref}
\usepackage{verbatim}% Long comments

\begin{document}

\title{Superconductivity and spin density wave in AA stacked bilayer
graphene}

\author{A.O. Sboychakov}
\affiliation{Institute for Theoretical and Applied Electrodynamics, Russian
Academy of Sciences, 125412 Moscow, Russia}

\author{A.L. Rakhmanov}
\affiliation{Institute for Theoretical and Applied Electrodynamics, Russian
Academy of Sciences, 125412 Moscow, Russia}

\author{A.V. Rozhkov }
\affiliation{Institute for Theoretical and Applied Electrodynamics, Russian
Academy of Sciences, 125412 Moscow, Russia}

\begin{abstract}
This work theoretically analyzes electronic ordering in AA-stacked bilayer
graphene and the role of the Coulomb interaction in these many-body
phenomena. Using the random phase approximation to account for screening,
we find intra-layer effective interactions to be much stronger than
inter-layer interactions; under certain circumstances, the latter may also
become attractive. At zero doping, the Coulomb repulsion stabilizes the
spin-density wave state, with a N{\'e}el temperature in the tens of Kelvin.
While dominant in the undoped system, the spin-density wave is destroyed by
sufficiently strong doping and a superconducting phase emerges. We find
that the effective Coulomb inter-layer interaction can give rise to
superconductivity. However, the corresponding critical temperature is
negligibly small, and phonon-mediated attraction must be introduced to
observe it. Strong intra-layer repulsion suppresses order parameters that
couple two intra-layer electrons. We point out a possible superconducting
state with finite Cooper pair momentum.
\end{abstract}

%\begin{keyword}
%bilayer graphene \sep AA-stacking \sep Coulomb interaction \sep RPA \sep
%SDW \sep superconductivity \sep phonons
%\end{keyword}

%\pacs{73.22.Pr, 73.22.Gk}
% 73.22.Pr --  Electronic structure of graphene
% 73.22.Gk -- Broken symmetry phases

\date{\today}
\maketitle

\section{Introduction}
Three different forms of bilayer 
graphene~\cite{CastrNrev,Abergel2010review_theor,bilayer_review2016}
are commonly discussed in the literature: AA-stacking (AA-BLG) or the
simple graphene bilayer, AB-stacking (AB-BLG) or Bernal phase, and the
twisted bilayer stacking (tBLG), which possess different band structures
and, consequently, different electronic properties. A striking
characteristic of the AA-BLG is an ideal nesting of the Fermi surface,
which has led to several intriguing theoretical predictions. In particular,
the nesting strongly enhances the effects of the electron-electron coupling
resulting in a stabilization of several symmetry-breaking phases, such as 
spin~\cite{aa_graph_prl2012,PrbROur,aa_graph_pha_sep_sboycha2013,
SDWandCDWAABLG}
and charge density 
waves~\cite{BreyAABLG,SDWandCDWAABLG},
and fractional 
metallicity~\cite{FracMetal,HalfMetalReviewJETPL2020},
among
others~\cite{su4aa2023prb}.
This makes AA-BLG an interesting object of a theoretical study that
attracted significant theoretical 
attention~\cite{Birowska2011,popov2013aa_intercalated_kr,
tabert2012aa_optics_theory,qdotAA2020,lu2023aa_andreev_theory}
despite the limited availability of high-quality 
samples~\cite{aa_graphite_lee2008,DiffStackingsBLG,borysiuk2011_aa,
iijima2009aa_graphene, caffrey2016aa_li_intercalated,
grubisic2023AA_exper}.
Moreover, tBLG twisted at a magic angle can be understood as a composition
of large regions of AA-BLG and AB-BLG, with doped electrons mostly
localized in the AA 
regions~\cite{bilayer_review2016,LowAngletBLGExp2015}.
We believe that the study of the electronic properties of the AA-BLG can
shed light on the peculiar features of the magic-angle tBLG, including its non-superconducting and superconducting ordered
phases~\cite{twist_exp_insul2018,Cao2018twisted_supercond,
cao2021twisted_supercond}.
Motivated by these
considerations, we analyze here many-body instabilities (both
superconducting and not) in AA-BLG at various levels of doping.

At zero-doping, our system displays ideal Fermi-surface nesting and is therefore dominated by insulating 
phases~\cite{aa_graph_prl2012,PrbROur,aa_graph_pha_sep_sboycha2013,
FracMetal, SDWandCDWAABLG, BreyAABLG} 
such as spin-density wave (SDW). In contrast, sufficiently strong doping
destroys the nesting and weakens the SDW, making superconducting order
possible.

As for specific mechanisms driving both instabilities, our investigation
reveals a peculiar role of the Coulomb interaction in these matters. For
the SDW, the influence of the Coulomb force is straightforward: the
stronger the repulsion, the larger the order parameter. The overall impact
of the Coulomb interaction on the superconductivity is more intricate. On
the one hand, in the classical BCS formulation the repulsion between the
electrons constituting Cooper pairs depletes the superconductivity. On the
other hand, we have the Kohn-Luttinger 
mechanism~\cite{kohn_lutt1965prl, kohn_lutt_graphene2014ZhETF}
that exclusively 
relies on the
Coulomb repulsion as the only source of the ``glue'' that binds the Cooper pairs
together. In other words, the Coulomb interaction can relate to the superconducting instability in a very non-trivial manner.

These considerations serve as a starting point of our analysis, which we further develop using the following three-stage program: 
(i)~For AA-BLG, we evaluate the screened Coulomb interaction within the random phase approximation (RPA). 
(ii)~Next, using the RPA interaction, we recover the SDW instability previously discussed mostly in the framework of the Hubbard model. 
(iii)~Finally, assuming that the sample is sufficiently doped to destroy the SDW, we explore a BCS-like superconducting mechanism for AA-BLG that incorporates both the phonon-mediated attraction and the screened Coulomb interaction.

Regarding point (iii), we emphasize that for AA-BLG the inter-layer effective interaction is weak, and can even be attractive. This RPA feature strongly favors the Cooper pairs whose constituent electrons are spatially segregated between the two layers of the bilayer. This observation significantly shortens the list of possible superconducting orders. In principle, the inter-layer effective interaction can itself lead to superconductivity. However, due to the weakness of the Coulomb-driven coupling constant, the corresponding order parameter turns out to be very small, and phonons must be included in the formalism.

A separate and very complicated issue, which we only touch on qualitatively, is the nature of the AA-BLG ground state at intermediate dopings, when the SDW is not completely destroyed. Depending on various model details, doped SDW may demonstrate phase separation partially frustrated by the long-range Coulomb interaction. Alternatively, the so-called fractional metallic states may also be realized. There are theoretical arguments and (indirect) experimental evidence for these phases in graphene-based systems. Both of these options are compatible with superconductivity that perhaps coexists with non-trivial spin texture.

In our analysis, we apply the mean-field approximation to consider
superconductivity and SDW order. In the case of the superconductivity we
use a familiar BCS approach, which has proven to be a reliable tool for
describing this phenomenon. The SDW order in our case arises due to nesting
of the Fermi surface. This result is common and is not a consequence of the
mean-field theory. For the graphene-based systems the applicability of the
mean-field approach was discussed in
Refs.~\onlinecite{PrbROur,ab_supercond2023sboychakov}. 
According these considerations a continuous transition, expected within the
mean-field framework, is replaced by a smooth crossover due to sensitivity
of two-dimensional systems to fluctuations. However, the crossover
temperature is of the same order as the mean field transition
temperatures.

The paper is organized as follows. In
Sec.~\ref{Model}
the studied model is formulated.
Section~\ref{polarOp}
is dedicated to the derivation of the screened Coulomb interaction in the
RPA framework. In
Sec.~\ref{SDW}
we consider the SDW state in the AA-BLG. In 
Sec.~\ref{BCS}
we analyze the superconductivity. Our results are discussed in
Sec.~\ref{Discussion}.
Additional calculations related to the superconductivity mechanism are
relegated to Appendices.
\section{The model}
%%%%%%%%%%%%%%%%%%%%%%%%%%%%%%%%%%%%%%%%%%%%%%%%%%%
\label{Model}
%%%%%%%%%%%%%%%%%%%%%%%%%%%%%%%%%%%%%%%%%%%%%%%%%%%

In AA-BLG carbon atoms in each sublattice of the top layer are located
right above the atoms of the same sublattice of the lower
layer~\cite{bilayer_review2016}.
We will take into account only intra-layer and inter-layer nearest neighbor
hoppings. Under this assumption the Hamiltonian of the system reads
\begin{eqnarray}
%%%%%%%%%%%%%%%%%%%%%%%%%%%%%%%%%%%%%%%%%%%%%%%%%%
\label{H0+Hi}
%%%%%%%%%%%%%%%%%%%%%%%%%%%%%%%%%%%%%%%%%%%%%%%%%%
  \!\!\!H
&=&
H_0+H_{\textrm{int}}
-
\mu\sum_{\mathbf{k}i\alpha\sigma}
	d^\dag_{\mathbf{k}i\alpha\sigma}
	d^{\phantom{\dag}}_{\mathbf{k}i\alpha\sigma},
\\
%%%%%%%%%%%%%%%%%%%%%%%%%%%%%%%%%%%%%%%%%%%%%%%%%%
\label{H0}
%%%%%%%%%%%%%%%%%%%%%%%%%%%%%%%%%%%%%%%%%%%%%%%%%%
  \!\!\!H_0
&=&
-\sum_{\mathbf{k}i\alpha\sigma}\!\!
	\left(t^{\alpha}_{\mathbf{k}}\;
		d^\dag_{\mathbf{k}i\alpha\sigma}
		d^{\phantom{\dag}}_{\mathbf{k}i\bar{\alpha}\sigma}
\!+\!
	t_0\;d^\dag_{\mathbf{k}i\alpha\sigma}
	d^{\phantom{\dag}}_{\mathbf{k}\bar{i}\alpha\sigma}
\right),
\\
%%%%%%%%%%%%%%%%%%%%%%%%%%%%%%%%%%%%%%%%%%%%%%%%%%
\label{Hint}
%%%%%%%%%%%%%%%%%%%%%%%%%%%%%%%%%%%%%%%%%%%%%%%%%%
\!\!\!H_{\textrm{int}}\!\!
&=&\!\!
  \frac{1}{2\cal N}\!\!\!
\sum_{\mathbf{kk}'\mathbf{p}\sigma\sigma'\atop ij\alpha\beta} \!\!\!
	d^{\dag}_{\mathbf{k}+\mathbf{p}i\alpha\sigma}
	d^{\phantom{\dag}}_{\mathbf{k}i\alpha\sigma}
	V^{ij}_{\mathbf{p}}	
	d^{\dag}_{\mathbf{k}'-\mathbf{p}j\beta\sigma'}
	d^{\phantom{\dag}}_{\mathbf{k}'j\beta\sigma'}.
\end{eqnarray}
Here
$H_0$
is the single-particle tight-binding Hamiltonian, the term
$H_{\textrm{int}}$
describes electron-electron interaction, $\mu$ is the chemical potential,
$\cal N$
is the number of unit cells in each layer, operator
$d^\dag_{\mathbf{k}i \alpha \sigma}$
(operator
$d^{\phantom{\dag}}_{\mathbf{k}i \alpha \sigma}$)
creates (annihilates) an electron with quasi-momentum
${\bf k}$
and spin projection $\sigma$ in layer $i$
[$i=0$
($i=1$)
corresponds to upper (lower) layer] on sublattice $\alpha$
[$\alpha=0$
($\alpha=1$)
represents sublattice $A$ (sublattice $B$)]. A bar over index $\alpha$
inverts the value of $\alpha$. The same convention applies to the layer
index $i$ in
Eq.~(\ref{H0}),
as well as to other binary-valued indices in expressions below. Further,
$t_0=0.36$\,eV
in
Eq.~\eqref{H0}
is the inter-layer nearest-neighbor hopping amplitude,
$t^{\alpha}_{\mathbf{k}}$
are defined according to
\begin{equation}
%%%%%%%%%%%%%%%%%%%%%%%%%%%%%%%%%%%%%%%%%%%%%%%%%%
\label{tA}
%%%%%%%%%%%%%%%%%%%%%%%%%%%%%%%%%%%%%%%%%%%%%%%%%%
t^{A}_{\mathbf{k}}=tf_{\mathbf{k}}\,,\;\;t^{B}_{\mathbf{k}}=tf^{*}_{\mathbf{k}}\,,
\end{equation}
where
$t=2.57$\,eV
is the intra-layer nearest-neighbor hopping amplitude, and function
$f_{\bf k}$
equals
\begin{equation}
%%%%%%%%%%%%%%%%%%%%%%%%%%%%%%%%%%%%%%%%%%%%%%%%%%
\label{f(k)}
%%%%%%%%%%%%%%%%%%%%%%%%%%%%%%%%%%%%%%%%%%%%%%%%%%
f_\mathbf{k}=1+2\exp{\left(-\sqrt{3}ik_xa/2\right)}\cos{(k_ya/2)}\,.
\end{equation}
In this definition
$a=2.46$\,\AA\
is the lattice constant.

The function
$V^{ij}_{\mathbf{p}}$
in
$H_{\rm int}$
encodes interaction of an electron in the layers $i$ with another electron
in the layer $j$, with transferred momentum
$\mathbf{p}$.
Below, we will consider an electron-electron coupling arising either due to
Coulomb repulsion or due to BCS-like electron-phonon scattering. The
interaction in our model is independent of the spin variables since the
model Hamiltonian does not include spin operators explicitly. Likewise,
$V^{ij}_{\mathbf{p}}$
is assumed to be independent of the sublattice indices, which is valid
provided that the interaction potential is sufficiently long-range.

\begin{figure}[t]
\centering
\includegraphics[width=0.48\textwidth]{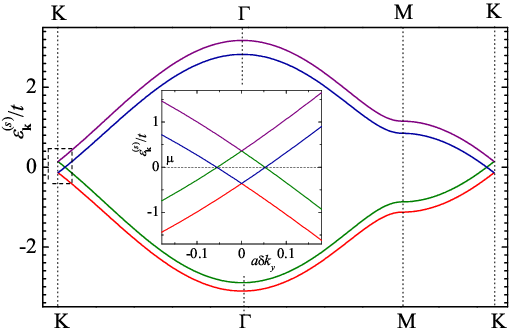}
\caption{The single-particle band structure of AA-BLG. Inset shows the
spectrum near a Dirac point (here,
$\mathbf{k}=\mathbf{K}+\delta k\mathbf{e}_{y}$).
Bands
$S = 2$ 
and 
$S=3$ 
cross zero energy, which corresponds to the Fermi level of the undoped
system.
%%%%%%%%%%%%%%%%%%%%%%%%%%%%%%%%%%%%%%%%%%%%%%%%%% 
\label{FigSpec}
%%%%%%%%%%%%%%%%%%%%%%%%%%%%%%%%%%%%%%%%%%%%%%%%%% 
}
\end{figure}

Diagonalizing
$H_0$
one finds four single-electron bands of AA-BLG. Their energies are
\begin{eqnarray}
\varepsilon^{(1)}_{\mathbf{k}}\!\!&=&\!\!-t_0-t\zeta_\mathbf{k}\,,\quad
\varepsilon^{(2)}_{\mathbf{k}}= -t_0+t\zeta_\mathbf{k}\,,\nonumber\\
\varepsilon^{(3)}_{\mathbf{k}} \!\!&=&\!\! +t_0-t\zeta_\mathbf{k}\,,\quad
\varepsilon^{(4)}_{\mathbf{k}}=+t_0+t\zeta_\mathbf{k}\,,
%%%%%%%%%%%%%%%%%%%%%%%%%%%%%%%%%%%%%%%%%%%%%%%%%%
\label{eigen_En_Fn}
%%%%%%%%%%%%%%%%%%%%%%%%%%%%%%%%%%%%%%%%%%%%%%%%%%
\end{eqnarray}
where
$\zeta_\mathbf{k}=|f_\mathbf{k}|$.
In the basis of the corresponding band operators
$\gamma_{\mathbf{k}S\sigma}$
the Hamiltonian
$H_0$
has a diagonal form
\begin{equation}
%%%%%%%%%%%%%%%%%%%%%%%%%%%%%%%%%%%%%%%%%%%%%%%%%%
\label{HamDiag}
%%%%%%%%%%%%%%%%%%%%%%%%%%%%%%%%%%%%%%%%%%%%%%%%%%
H_0
=
\sum_{\mathbf{k}S\sigma}
	\varepsilon^{(S)}_{\mathbf{k}}
	\gamma^\dag_{\mathbf{k}S\sigma}
	\gamma^{\phantom{\dag}}_{\mathbf{k}S\sigma}\,,
\end{equation}
where index
$S=1,\ldots,\,4$
The AA-BLG band structure is shown in Fig.~\ref{FigSpec}.

The bands
$S = 2$
and
$S = 3$
cross the Fermi level forming a Fermi surface. The Fermi surface for band~2
(band~3) consists of two approximately circular electron (hole) sheets.
Each sheet is centered at one of two non-equivalent Dirac points
$\mathbf{K}_{1}=4\pi(0,\,1)/(3a)$
and
$\mathbf{K}_{-1}=-\mathbf{K}_1$.
The Fermi momenta are
\begin{eqnarray}
%%%%%%%%%%%%%%%%%%%%%%%%%%%%%%%%%%%%%%%%%%%%%%%%%%
\label{kF2}
%%%%%%%%%%%%%%%%%%%%%%%%%%%%%%%%%%%%%%%%%%%%%%%%%%
\text{band 2 (electrons):}
\quad
k_{\rm F +}= k_{\rm F} + \mu/v_{\rm F},
\\
%%%%%%%%%%%%%%%%%%%%%%%%%%%%%%%%%%%%%%%%%%%%%%%%%%
\label{kF3}
%%%%%%%%%%%%%%%%%%%%%%%%%%%%%%%%%%%%%%%%%%%%%%%%%%
\text{band 3 (holes):}
\quad
k_{\rm F -}= k_{\rm F} - \mu/v_{\rm F},
\end{eqnarray}
where
$k_{\rm F} = t_0/v_{\rm F}$
is the Fermi momentum of undoped AA-BLG, and
$v_{\rm F}=\sqrt{3}ta/2$
is the graphene Fermi velocity.

When
$\mu=0$,
which corresponds to the undoped bilayer, Fermi momenta of the bands~2
(electrons) and band~3 (holes) coincide. In other words, we have perfect
Fermi surface nesting with nesting vectors
$\mathbf{0}$
and
$\mathbf{Q}_0=\mathbf{K}_{1}-\mathbf{K}_{-1}$.
This feature is very robust: even if more distant intra-layer and
inter-layer hopping terms are introduced into
$H_0$,
the Fermi surface sheets remain
coincident~\cite{aa_graph_prl2012}.
Such a Fermi liquid is unstable with respect to the spontaneous symmetry
breaking.

Here we are interested in the low-doped samples
($|\mu|\ll t_0$)
and low-energy states. Thus, in the simplest approximation, we can omit the
bands
$S = 1,\,4$
from the formalism. In this approximation the relation between
$d_{\mathbf{k}i\alpha\sigma}$
and
$\gamma_{\mathbf{k}S\sigma}$
can be expressed as
\begin{eqnarray}
%%%%%%%%%%%%%%%%%%%%%%%%%%%%%%%%%%%%%%%%%%%%%%%%%%
\label{eigen_En_Fn_1}
%%%%%%%%%%%%%%%%%%%%%%%%%%%%%%%%%%%%%%%%%%%%%%%%%%
d_{\mathbf{k}i\alpha\sigma}
=
\frac{1}{2}
\exp\!\! \left[ \frac{i}{2} (-1)^\alpha \varphi_\mathbf{k} \right]
\!\left[
	(-1)^{\alpha}\gamma_{\mathbf{k}2\sigma}
	+
	(-1)^i\gamma_{\mathbf{k}3\sigma}
\right]\!,
\qquad
\end{eqnarray}
where
$\varphi_\mathbf{k}=\textrm{arg}(f_\mathbf{k})$.
Additionally, we approximate
$f_\mathbf{k}$
near two Dirac points according to
\begin{equation}
%%%%%%%%%%%%%%%%%%%%%%%%%%%%%%%%%%%%%%%%%%%%%%%%%%
\label{f_Smalk}
%%%%%%%%%%%%%%%%%%%%%%%%%%%%%%%%%%%%%%%%%%%%%%%%%%
f_\mathbf{\mathbf{K}_{\xi}+\mathbf{q}}
\approx
v_{\rm F} (iq_x - \xi q_y),
\quad
\xi = \pm 1.
\end{equation}
Consequently, near
$\mathbf{K}_{\xi}$
we have
$t\zeta_{\mathbf{K}_{\xi}+\mathbf{q}}=v_{\rm F}q$
and
$\varphi_{\mathbf{K}_{\xi}+\mathbf{q}}
=
\frac{\pi}{2} + \xi\phi_{\mathbf{q}}$,
where
$\phi_{\mathbf{q}}$
is the polar angle corresponding to vector
$\mathbf{q}$,
and
$q = |{\bf q}|$.
From
Eqs.~\eqref{eigen_En_Fn}
we obtain
\begin{eqnarray}
%%%%%%%%%%%%%%%%%%%%%%%%%%%%%%%%%%%%%%%%%%%%%%%%%%
\label{eigen_En_Fn_2}
%%%%%%%%%%%%%%%%%%%%%%%%%%%%%%%%%%%%%%%%%%%%%%%%%%
\varepsilon^{(2,3)}_{\mathbf{K}_{\xi}+\mathbf{q}}
&\approx&
\pm\left(v_{\rm F}q-t_0\right)\equiv\pm\varepsilon_{\mathbf{q}}.
\end{eqnarray}
This relation approximates the low-energy electronic spectrum of AA-BLG. In
the doped sample
($\mu\neq 0$),
the Fermi surfaces for the electronic band~2 and hole band~3 become
non-identical. This destroys perfect nesting both at nesting vector
$\mathbf{0}$
and
$\mathbf{Q}_0=\mathbf{K}_{1}-\mathbf{K}_{-1}$.
This property has important implications for stability of the SDW.

\begin{figure*}[t]
\centering
\includegraphics[width=0.32\textwidth]{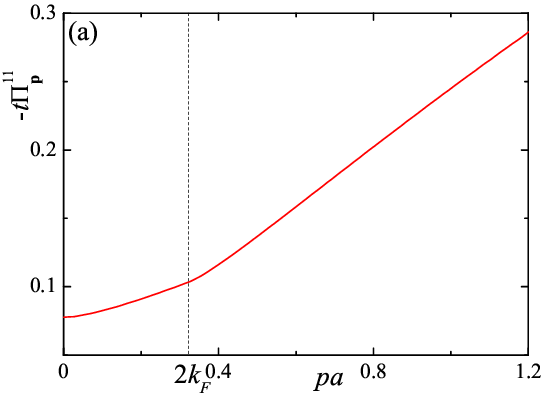}
\includegraphics[width=0.32\textwidth]{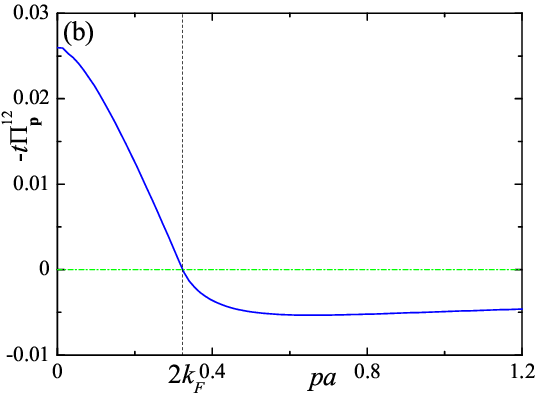}
\includegraphics[width=0.32\textwidth]{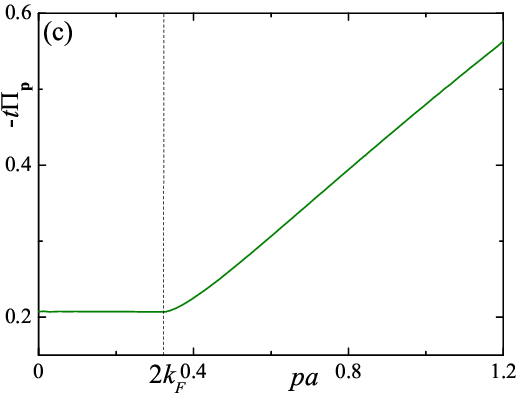}
\caption{
%%%%%%%%%%%%%%%%%%%%%%%%%%%%%%%%%%%%%%%%%%%%%%%%%%
\label{FigP}
%%%%%%%%%%%%%%%%%%%%%%%%%%%%%%%%%%%%%%%%%%%%%%%%%%
The dependencies of
$\Pi^{11}_{\mathbf{p}}$
(a),
$\Pi^{12}_{\mathbf{p}}$
(b), and
$\Pi_{\mathbf{p}}=\sum_{ij}\Pi^{ij}_{\mathbf{p}}$
(c) on momentum
${\bf p}$
for
$\mathbf{p}=p(1,\,0)$,
calculated at
$\mu=0$.
Other parameters of the model are presented in the text.
}
\end{figure*}

\section{Polarization operator and screened Coulomb interaction}
%%%%%%%%%%%%%%%%%%%%%%%%%%%%%%%%%%%%%%%%%%%%%%%%%%
\label{polarOp}
%%%%%%%%%%%%%%%%%%%%%%%%%%%%%%%%%%%%%%%%%%%%%%%%%%

In this section we derive the screened Coulomb interaction in AA-BLG within
the RPA. The latter approximation is exact in the limit of large degeneracy
factor
$N_f$.
For a typical metal,
$N_f = 2$,
which accounts for the spin degeneracy. In graphene and graphene-based
systems
$N_f$
is enhanced and becomes
$N_f = 4$
due to additional valley degeneracy. We view this as an indication of
applicability of the RPA scheme for AA-BLG.

To execute the RPA calculation it is convenient to express the bare Coulomb
interaction as a
$2\times2$
matrix function
\begin{equation}
%%%%%%%%%%%%%%%%%%%%%%%%%%%%%%%%%%%%%%%%%%%%%%%%%%
\label{V0}
%%%%%%%%%%%%%%%%%%%%%%%%%%%%%%%%%%%%%%%%%%%%%%%%%%
\hat{V}^{(0)}_{\mathbf{p}}=\frac{A}{p}\left(\begin{array}{cc}
1&e^{-pd}\\
e^{-pd}&1
\end{array}\right),\;\;A=\frac{2\pi e^2}{v_c\epsilon}\,,
\end{equation}
where
$d=3.35$\,\AA\ is the inter-layer distance,
$v_c=\sqrt{3}a^2/2$
is the area of the graphene unit cell, and $\epsilon$ is the dielectric
constant of the media surrounding the graphene sample. More precisely, we
assume that the sample is placed on the dielectric substrate with
dielectric constant 
$\epsilon_s$. 
If the sample is covered by a substance with the same
$\epsilon_s$,
then 
$\epsilon=\epsilon_s$.
If the sample is not covered, then 
$\epsilon=(1+\epsilon_s)/2$.
For suspended sample 
$\epsilon=1$. 
Matrix elements
$V^{(0)11}_{\mathbf{p}}$
and
$V^{(0)22}_{\mathbf{p}}$
represent intra-layer interactions, while
$V^{(0)12}_{\mathbf{p}}$
and
$V^{(0)21}_{\mathbf{p}}$
represent the interaction between electrons in different layers.

To find renormalized Coulomb interaction we calculate the static
polarization operator
$\hat{\Pi}_{\bf p}$,
which is also a
$2\times2$
matrix function of the transferred momentum
$\mathbf{p}$.
Its matrix elements can be written in the
form~\cite{Pol6FourBandRossiPRB2012}
\begin{eqnarray}
%%%%%%%%%%%%%%%%%%%%%%%%%%%%%%%%%%%%%%%%%%%%%%%%%%
\label{P}
%%%%%%%%%%%%%%%%%%%%%%%%%%%%%%%%%%%%%%%%%%%%%%%%%%
\Pi^{ij}_{\mathbf{p}}
=
2\sum_{SS'}\int\!\frac{d^2\mathbf{k}}{v_{\rm BZ}}
\frac{
	n_{\rm F}\left(\varepsilon^{(S)}_{\mathbf{k}}\right)
	-
	n_{\rm F}\left(\varepsilon^{(S')}_{\mathbf{k}+\mathbf{p}}\right)
}
{
	\varepsilon^{(S)}_{\mathbf{k}}
	-
	\varepsilon^{(S')}_{\mathbf{k}+\mathbf{p}}
}\times
\nonumber
\\
\Big(
	\sum_{\alpha}
		\Phi^{(S)}_{\mathbf{k}i\alpha}
		\Phi^{(S')*}_{\mathbf{k}+\mathbf{p}i\alpha}
\Big)
\Big(
	\sum_{\beta}
		\Phi^{(S)*}_{\mathbf{k}j\beta}
		\Phi^{(S')}_{\mathbf{k}+\mathbf{p}j\beta}
\Big),
\end{eqnarray}
where
$\Phi^{(S)}_{\mathbf{k}i\alpha}$
are the eigenfunctions of the single-particle
Hamiltonian~\eqref{H0},
$v_{\rm BZ}=4\pi^2/v_c$
is the Brillouin zone area, and
\begin{equation}
%%%%%%%%%%%%%%%%%%%%%%%%%%%%%%%%%%%%%%%%%%%%%%%%%%%
\label{nF}
%%%%%%%%%%%%%%%%%%%%%%%%%%%%%%%%%%%%%%%%%%%%%%%%%%%
n_{\rm F} (E)=\frac{1}{e^{(E-\mu)/T}+1}
\end{equation}
is the Fermi function. We calculate
$\Pi^{ij}_{\mathbf{p}}$
numerically at
$T=10^{-4}t$.
The result of calculation at zero doping,
$\mu=0$,
is shown in
Fig.~\ref{FigP}. Since layers are identical, we have
$\Pi^{11}_{\mathbf{p}}=\Pi^{22}_{\mathbf{p}}$.
Also
$\Pi^{12}_{\mathbf{p}}=\Pi^{21}_{\mathbf{p}}$.

At not too large $p$ the polarization operator is virtually insensitive to
the direction of
$\mathbf{p}$.
The value of
$\Pi^{11}_{\mathbf{p}}$
is always negative, while
$\Pi^{12}_{\mathbf{p}}$
is negative at small $p$ and positive at larger $p$, changing sign at
$p \approx 2 k_{\rm F}$.
As for
$-\Pi^{11}_{\mathbf{p}}$,
it increases when $p$ grows. This increase is linear in $p$ for $p$ above
$2 k_{\rm F}$.
We also clearly see Kohn anomaly in
$\Pi^{11}_{\mathbf{p}}$
at
$p=2k_{\rm F}$.
The total polarization operator
$\Pi_{\mathbf{p}}=\sum_{ij}\Pi^{ij}_{\mathbf{p}}$
is constant for
$p<2k_{\rm F}$
and
$-\Pi_{\mathbf{p}}$
increases linearly at larger $p$, similar to what is calculated for
single-layer graphene. Such a behavior of
$\Pi_{\mathbf{p}}$
is consistent with that obtained in
Ref.~\onlinecite{BreyPAA}
in the framework of continuum approximation.

The renormalized Coulomb interaction can be expressed in the matrix form as
\begin{equation}
%%%%%%%%%%%%%%%%%%%%%%%%%%%%%%%%%%%%%%%%%%%%%%%%%%%
\label{VRPA}
%%%%%%%%%%%%%%%%%%%%%%%%%%%%%%%%%%%%%%%%%%%%%%%%%%%
\hat{V}_{\mathbf{p}}^{\vphantom{(0)}}
=
\hat{V}^{(0)}_{\mathbf{p}}
\left[
	1
	-
	\hat{\Pi}_{\mathbf{p}}^{\vphantom{(0)}} \hat{V}^{(0)}_{\mathbf{p}}
\right]^{-1}.
\end{equation}
Performing the matrix inversion, one obtains
\begin{widetext}
\begin{eqnarray}
%%%%%%%%%%%%%%%%%%%%%%%%%%%%%%%%%%%%%%%%%%%%%%%%%%
\label{V11}
%%%%%%%%%%%%%%%%%%%%%%%%%%%%%%%%%%%%%%%%%%%%%%%%%%
V^{11}_{\mathbf{p}}
=
V^{22}_{\mathbf{p}}
=
A\frac{1-\frac{A}{p}(1-e^{-2pd})\Pi^{11}_{\mathbf{p}}}
{
	p-2A(\Pi^{11}_{\mathbf{p}}+e^{-pd}\Pi^{12}_{\mathbf{p}})+
	\frac{A^2}{p}(1-e^{-2pd})
	[(\Pi^{11}_{\mathbf{p}})^2-(\Pi^{12}_{\mathbf{p}})^2]
},
\\
%%%%%%%%%%%%%%%%%%%%%%%%%%%%%%%%%%%%%%%%%%%%%%%%%%
\label{V12}
%%%%%%%%%%%%%%%%%%%%%%%%%%%%%%%%%%%%%%%%%%%%%%%%%%
V^{12}_{\mathbf{p}}
=
V^{21}_{\mathbf{p}}
=
A\frac{e^{-pd}+\frac{A}{p}(1-e^{-2pd})\Pi^{12}_{\mathbf{p}}}
{
	p-2A(\Pi^{11}_{\mathbf{p}}+e^{-pd}\Pi^{12}_{\mathbf{p}})+
	\frac{A^2}{p}(1-e^{-2pd})
	[(\Pi^{11}_{\mathbf{p}})^2-(\Pi^{12}_{\mathbf{p}})^2]
}.
\end{eqnarray}
\end{widetext}
%\section{Superconductivity and Coulomb electron-electron coupling}
%%%%%%%%%%%%%%%%%%%%%%%%%%%%%%%%%%%%%%%%%%%%%%%%%%
%\label{CoulombSC}
%%%%%%%%%%%%%%%%%%%%%%%%%%%%%%%%%%%%%%%%%%%%%%%%%%
Related results can be found in the literature on the Coulomb drag in
two-dimensional systems; see, for example,
Refs.~\onlinecite{kamenev1995theory_prb,flensberg1995theory_prb,
coulomb_drag_graphene2012theory}.
Note, however, that our equations are more general than those derived to
describe the Coulomb drag effect. Indeed, in a typical Coulomb drag
experiment the layers are insulated from each other, thus, electron hopping
between them is zero, and, consequently,
$\Pi^{12}_{\mathbf{q}}\equiv0$.
We do not make such a simplification.

\begin{figure}
\centering
\includegraphics[width=0.95\columnwidth]{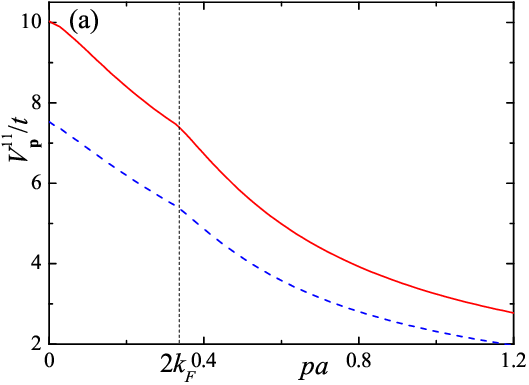}
\includegraphics[width=0.95\columnwidth]{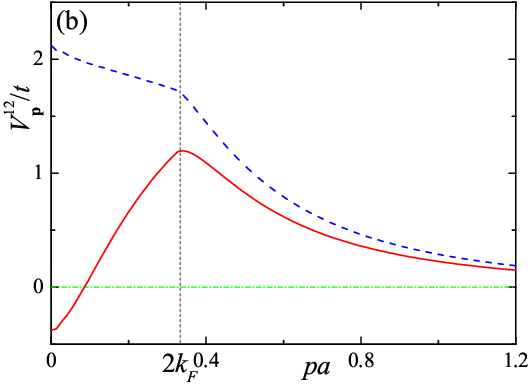}
\caption{
%%%%%%%%%%%%%%%%%%%%%%%%%%%%%%%%%%%%%%%%%%%%%%%%%%
\label{FigV}
%%%%%%%%%%%%%%%%%%%%%%%%%%%%%%%%%%%%%%%%%%%%%%%%%%
The dependencies of
$V^{11}_{\mathbf{p}}$
(a) and
$V^{12}_{\mathbf{p}}$
(b) on momentum
${\bf p}$
for
$\mathbf{p}=p(1,\,0)$
calculated at
$\epsilon=1$
(red solid curves) and at
$\epsilon=3$
(blue dashed curves), see
Eqs.~(\ref{V11})
and~(\ref{V12}).
For all curves
$\mu=0$.
Other parameters of the model are presented in the text.
}
\end{figure}

Renormalized interaction
$V^{ij}_{\mathbf{p}}$
evaluated according to these formulas is shown in
Fig.~\ref{FigV}
for
$\epsilon=1$
and
$\epsilon=3$.

Observe that, for
$p\lesssim 2k_{\rm F}$,
the value
$\Pi^{12}_{\mathbf{p}}$
is negative [see
Fig.~\ref{FigP}\,(b)].
Equation~(\ref{V12})
indicates that, if
$2 A d |\Pi_{\bf p}^{12}|$
exceeds unity at low
${\bf p}$,
then
$V^{12}_{\mathbf{p}}$
becomes attractive for small transferred momenta. Our calculations show
that this is indeed the case when $\epsilon$ is close to unity. In
Fig.~\ref{FigV}\,(b)
we see that for
$\epsilon=1$
the inter-layer component
$V^{12}_{\mathbf{p}}$
is negative at small $p$.

The latter finding seems to imply that a superconducting phase might be
stabilized by the screened Coulomb repulsion in AA-BLG. However, within the
framework of the Coulomb-only superconducting mechanism, our estimate for
the critical temperature
$T_{\rm sc}$
shows that this temperature is virtually zero even for the most favorable
case of
$\epsilon=1$
[e.g., in terms of the usual BCS superconductivity with
$T_{\rm sc}\propto \exp(-1/\lambda )$
our estimates corresponds to
$\lambda \approx0.004$, see~\ref{AppendixA}].
Thus, phonon-free superconducting mechanism in AA-BLG cannot be
rationalized within our framework. At the same time, our RPA investigation
offers certain suggestions about a possible structure of a phonon-based
superconducting order parameter, as we will see in
Sec.~\ref{BCS}.

\begin{figure}[t]
\centering
\includegraphics[width=0.42\textwidth]{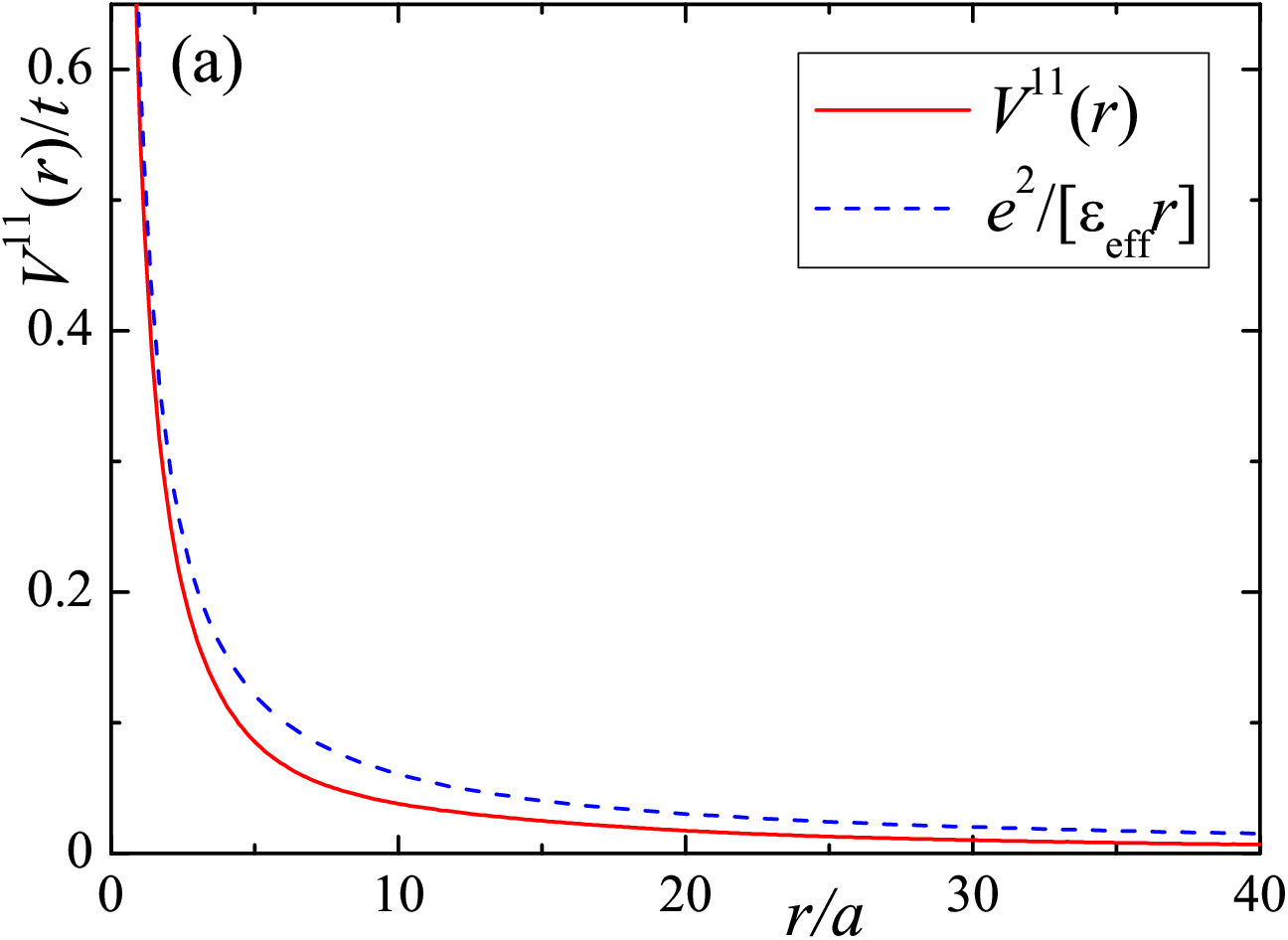}
\includegraphics[width=0.42\textwidth]{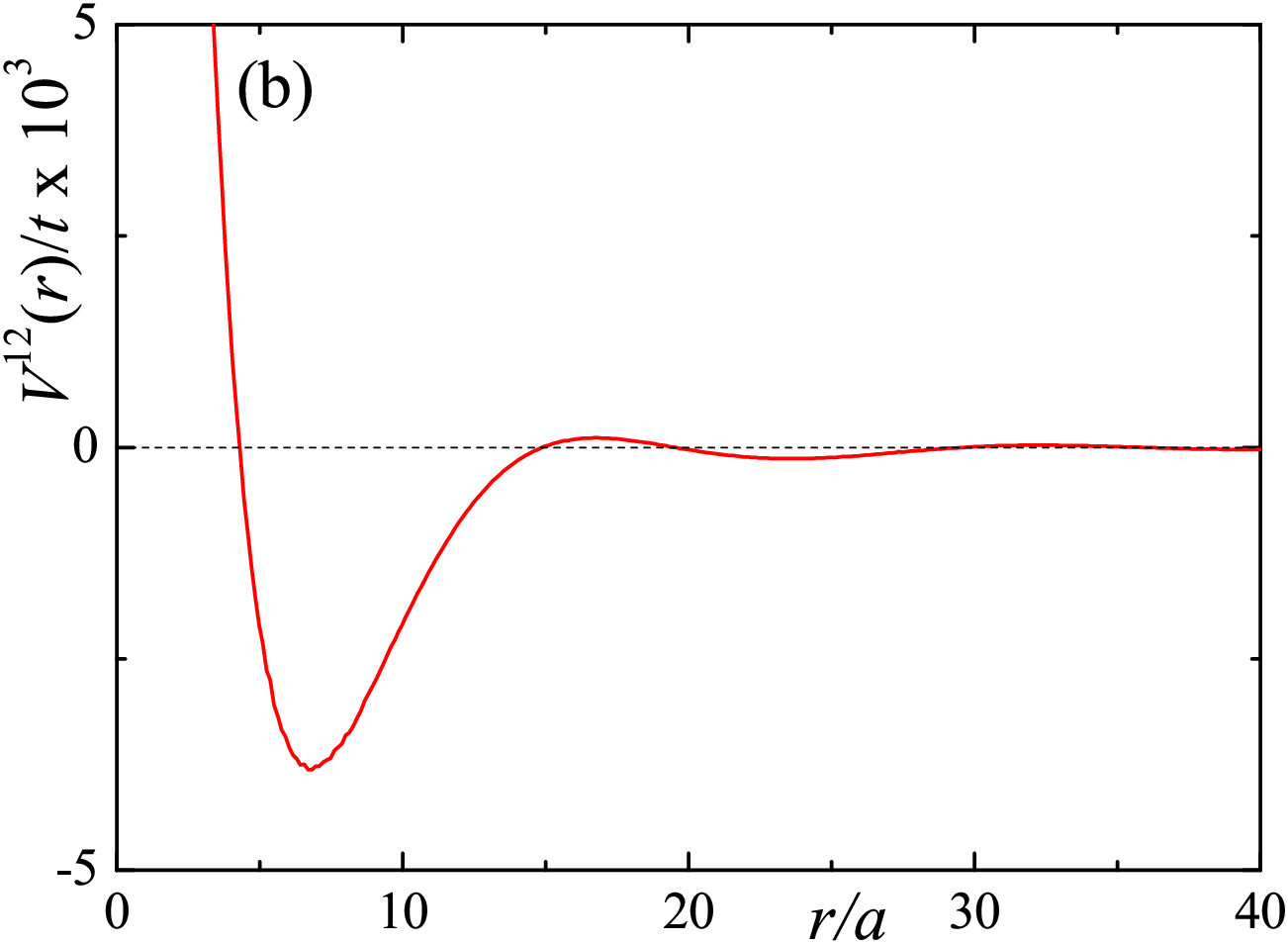}
\caption{The $r$ dependencies of the
renormalized intra-layer (a) and inter-layer (b) Coulomb potentials (red
solid curves). The Friedel oscillations are clearly seen for
$V^{12}(r)$.
Blue dashed curve in panel~(a) shows the dependence
$V_0(r)=e^2/(\varepsilon_{\text{eff}}r)$,
where
$\varepsilon_{\text{eff}}\approx\epsilon+\pi e^2/(2v_F)$
is effective dielectric constant. For all curves 
$\mu=0$. 
Other parameters of the model are presented in the text.
%%%%%%%%%%%%%%%%%%%%%%%%%%%%%%%%%%%%%%%%%%%%%%%%%%
\label{FigVR}
%%%%%%%%%%%%%%%%%%%%%%%%%%%%%%%%%%%%%%%%%%%%%%%%%%
}
\end{figure}

Knowing the functions 
$V^{ij}_{\mathbf{p}}$ 
one can perform a Fourier transform and calculate the renormalized Coulomb
interaction 
$V^{ij}(\mathbf{r})$
in real space. Since the polarization operator is virtually independent of
the direction of
$\mathbf{p}$, 
thus, to a good approximation, the functions
$V^{ij}(\mathbf{r})$ 
depend only on the absolute value of
$\mathbf{r}$.
Corresponding curves are shown in 
Fig.~\ref{FigVR}. 

In
Fig.~\ref{FigVR}(b) 
we clearly see Friedel oscillations emerging due the Kohn anomaly of the
polarization operator at
$p=2k_F$. 
The inter-layer Coulomb interaction is negative for some values of $r$,
which can lead to the superconducting instability driven by the Coulomb
interaction (see 
\ref{AppendixA} 
for estimation of the transition temperature).

At large $p$, the polarization operator 
behaves~\cite{note_on_PSLG}
as
$\Pi^{ij}_{\mathbf{p}}\approx\delta_{ij}\Pi^{\text{SLG}}_{\mathbf{p}}$,
where $\Pi^{\text{SLG}}_{\mathbf{p}}=-pv_c/(4v_F)$ is static polarization
operator of the undoped single-layer
graphene~\cite{InteractionInGrapheneReview2012}. As a result, at short
distances, the intra-layer Coulomb interaction behaves as
$e^2/(\varepsilon_{\text{eff}}r)\equiv V_0(r)$ [see blue dashed curve on
Fig.~\ref{FigVR}(a)], where $\varepsilon_{\text{eff}}\approx\epsilon+\pi
e^2/(2v_F)$ is RPA single-layer graphene dielectric constant. 

As for the intra-layer interaction
$V^{11}(\mathbf{r})$, 
it is always positive and smaller than 
$V_0(r)$ 
for any $r$. The curve
$V^{11}(\mathbf{r})$
does not demonstrate any discernible Friedel oscillations.

Finally, we would like to make the following comment. In the limit of
$d\to0$
we recover the widely used result
$V^{ij}_{\mathbf{p}} = A/(p - A\Pi_{\mathbf{p}})$.
We argued that this approximation may perform poorly in realistic
situations. For details, one can consult Appendix~B of
Ref.~\onlinecite{ab_supercond2023sboychakov}.

\section{Spin-density wave state}
%%%%%%%%%%%%%%%%%%%%%%%%%%%%%%%%%%%%%%%%%%%%%%%%%%
\label{SDW}
%%%%%%%%%%%%%%%%%%%%%%%%%%%%%%%%%%%%%%%%%%%%%%%%%%

In this section we develop a mean-field theory for the SDW phase at zero
temperature using the obtained above screened Coulomb interaction. The
order parameter for the SDW state couples the electrons in the band
$S=2$
and holes in the band
$S=3$
with opposite spin projections. We substitute operators
$d_{\mathbf{k}i\alpha\sigma}$
in the form of
Eq.~\eqref{eigen_En_Fn_1}
%Eq.~\eqref{d_xi}
in the interaction
Hamiltonian~\eqref{Hint}
and keep only the terms pertinent to the SDW pairing. As a result we obtain
\begin{eqnarray}
%%%%%%%%%%%%%%%%%%%%%%%%%%%%%%%%%%%%%%%%%%%%%%%%%%
\label{HintPsi}
%%%%%%%%%%%%%%%%%%%%%%%%%%%%%%%%%%%%%%%%%%%%%%%%%%
H_{\text{int}}
&=&
-\frac{1}{2\cal N}
\sum_{\mathbf{kk}'\sigma}
	\left(
		\gamma^{\dag}_{\mathbf{k}2\sigma}
		\gamma^{\phantom{dag}}_{\mathbf{k}3\bar{\sigma}}
		\Gamma^{(1)}_{\mathbf{kk}'}
		\gamma^{\dag}_{\mathbf{k}'3\bar{\sigma}}
		\gamma^{\phantom{dag}}_{\mathbf{k}'2\sigma}
	\right.
+
\nonumber\\
&&
	\left.
		\gamma^{\dag}_{\mathbf{k}2\sigma}
		\gamma^{\phantom{dag}}_{\mathbf{k}3\bar{\sigma}}
		\Gamma^{(2)}_{\mathbf{kk}'}
		\gamma^{\dag}_{\mathbf{k}'2\bar{\sigma}}
		\gamma^{\phantom{dag}}_{\mathbf{k}'3\sigma}
		+ {\rm H.c.}
	\right),
\end{eqnarray}
where
\begin{eqnarray}
%%%%%%%%%%%%%%%%%%%%%%%%%%%%%%%%%%%%%%%%%%%%%%%%%%
\label{Gamma12}
%%%%%%%%%%%%%%%%%%%%%%%%%%%%%%%%%%%%%%%%%%%%%%%%%%
\Gamma^{(1)}_{\mathbf{kk}'}
=
\sum_{ij}\Big(
	\sum_{\alpha}
		\Phi^{(2)*}_{\mathbf{k}i\alpha}
		\Phi^{(2)}_{\mathbf{k}'i\alpha}
	\Big)
	V^{ij}_{\mathbf{k}-\mathbf{k}'}
	\Big(
	\sum_{\beta}
		\Phi^{(3)}_{\mathbf{k}j\beta}
		\Phi^{(3)*}_{\mathbf{k}'j\beta}
	\Big),
\;\;\;\;
\nonumber\\
\Gamma^{(2)}_{\mathbf{kk}'}
=
\sum_{ij}\Big(
	\sum_{\alpha}
		\Phi^{(2)*}_{\mathbf{k}i\alpha}
		\Phi^{(3)}_{\mathbf{k}'i\alpha}\Big)
	V^{ij}_{\mathbf{k}-\mathbf{k}'}
	\Big(
	\sum_{\beta}
		\Phi^{(3)}_{\mathbf{k}j\beta}
		\Phi^{(2)*}_{\mathbf{k}'j\beta}\Big).
\;\;\;\;\;
\end{eqnarray}
We suppose that in the SDW state the following expectation values are
non-zero
\begin{equation}
%%%%%%%%%%%%%%%%%%%%%%%%%%%%%%%%%%%%%%%%%%%%%%%%%%
\label{etaSDW}
%%%%%%%%%%%%%%%%%%%%%%%%%%%%%%%%%%%%%%%%%%%%%%%%%%
\eta_{\mathbf{k}}
=
\left\langle
	\gamma^{\dag}_{\mathbf{k}3\bar{\sigma}}
	\gamma^{\phantom{dag}}_{\mathbf{k}2\sigma}
\right\rangle.
\end{equation}
They are additionally assumed to be independent of $\sigma$, which fixes
the SDW polarization direction to be collinear with the $x$-axis. Next, we
introduce the SDW order parameter as
\begin{equation}
%%%%%%%%%%%%%%%%%%%%%%%%%%%%%%%%%%%%%%%%%%%%%%%%%%%
\label{deltaSDW}
%%%%%%%%%%%%%%%%%%%%%%%%%%%%%%%%%%%%%%%%%%%%%%%%%%%
\Delta^{\text{SDW}}_{\mathbf{k}}
=
\frac{1}{\cal N}\sum_{\mathbf{k}'}
	\left(
		\Gamma^{(1)}_{\mathbf{kk}'}
		\eta_{\mathbf{k}'}^{\vphantom{\dagger}}
		+
		\Gamma^{(2)}_{\mathbf{kk}'}
		\eta^{*{\vphantom{\dagger}}}_{\mathbf{k}'}
\right)\,.
\end{equation}
Following the standard mean-field decoupling scheme in
Eq.~\eqref{HintPsi},
we obtain the mean-field Hamiltonian and calculate the grand potential
$\Omega$ of the system. Minimization of $\Omega$ allows us to derive the
self-consistency equation for the SDW order parameter
\begin{equation}
%%%%%%%%%%%%%%%%%%%%%%%%%%%%%%%%%%%%%%%%%%%%%%%%%%
\label{DeltaSDWeq}
%%%%%%%%%%%%%%%%%%%%%%%%%%%%%%%%%%%%%%%%%%%%%%%%%%
\Delta^{\text{SDW}}_{\mathbf{k}}
=
\int\!\frac{d^2\mathbf{k}'}{v_{\rm BZ}}\,
	\frac{
		\Gamma^{(1)}_{\mathbf{kk}'}
		\Delta^{\text{SDW}}_{\mathbf{k}'}
		+
		\Gamma^{(2)}_{\mathbf{kk}'}
		\Delta^{\text{SDW}*}_{\mathbf{k}'}
	}
	{
		2\sqrt{
			(t \zeta_{\mathbf{k}'}-t_0)^2
			+
			|\Delta^{\text{SDW}}_{\mathbf{k}'}|^2
		}
	}.
\end{equation}
To solve the integral
equation~\eqref{DeltaSDWeq}
we postulate that
$\Delta^{\text{SDW}}_{\mathbf{k}}$
is non-zero only in the region near each Dirac point. Moreover, we
use the following simplification
\begin{equation}
%%%%%%%%%%%%%%%%%%%%%%%%%%%%%%%%%%%%%%%%%%%%%%%%%%
\label{ansatz}
%%%%%%%%%%%%%%%%%%%%%%%%%%%%%%%%%%%%%%%%%%%%%%%%%%
\Delta^{\text{SDW}}_{\mathbf{K}_{\xi}+\mathbf{p}}=\left\{\begin{array}{cl}
\Delta^{\text{SDW}}_{\xi p}&,\;\;|\mathbf{p}|<K_0\\
0&,\;\;|\mathbf{p}|>K_0
\end{array}\right.\,,
\end{equation}
where the function
$\Delta^{\text{SDW}}_{\xi p}$
is assumed to be real and depends only on the absolute value of the vector
$\mathbf{p}$,
while the cutoff momentum
$K_0=2\pi/(3a)$
is chosen to guarantee that regions corresponding to the different
Dirac-points valleys do not intersect.

\begin{figure}
\centering
\includegraphics[width=0.75\columnwidth]{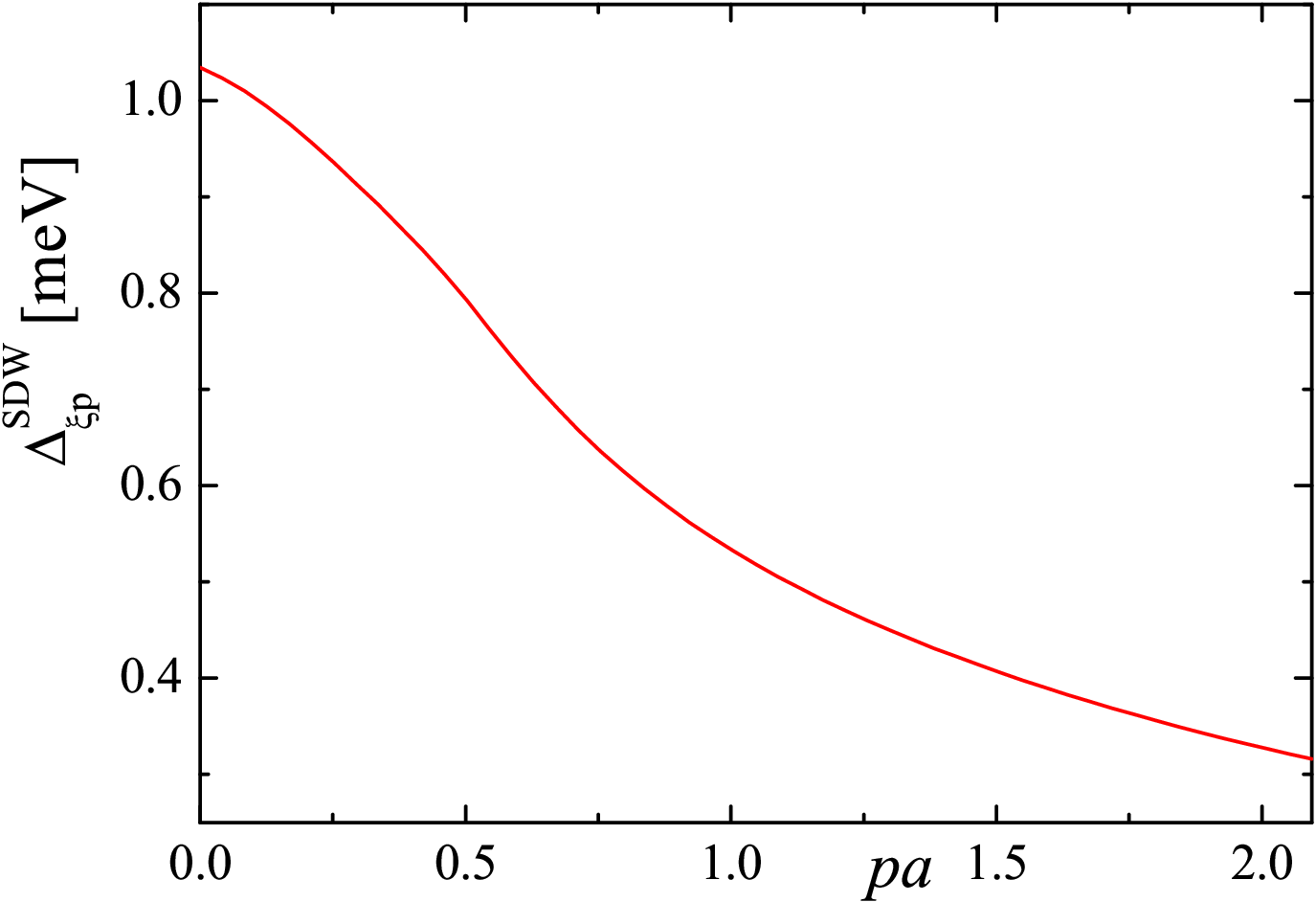}
\caption{
%%%%%%%%%%%%%%%%%%%%%%%%%%%%%%%%%%%%%%%%%%%%%%%%%%
\label{SDW_OP}
%%%%%%%%%%%%%%%%%%%%%%%%%%%%%%%%%%%%%%%%%%%%%%%%%%
The dependencies of
$\Delta^{\rm SDW}_{\xi p}$
versus $p$. The order parameter is calculated solving
Eq.~(\ref{DeltaSDWeq})
for
$\mu=0$
and
$\epsilon = 1$.
Other model parameters are presented in the text.
}
\end{figure}
Using the
ansatz~\eqref{ansatz},
we solve
Eq.~\eqref{DeltaSDWeq}
numerically by the successive iterations method. The result of this
calculations is shown in
Fig.~\ref{SDW_OP}.
We see that
$\Delta^{\text{SDW}}_{\xi p}$
is a decreasing function of $p$: at the edge of the region
$p<K_0$
it is about three times smaller than at
$p=0$.
For
$\epsilon=1$
the maximum value of
$\Delta^{\text{SDW}}_{\xi p}$
is
$\Delta^{\text{SDW}}_{\xi0}\approx 1$\,meV.
This corresponds to the transition temperature about
$10$\,K.
Note that in our model we do not include the on-site Hubbard term, which
will increase the above estimate.

Since doping spoils the ideal nesting, the SDW order weakens when extra
charges enter the system. As it was shown in
Refs.~\onlinecite{PrbROur,aa_graph_pha_sep_sboycha2013},
the commensurate SDW order disappears when the doping concentration per
site $x$ exceeds the critical value
$x_c \approx \Delta^{\text{SDW}}_{\xi0}t_0/(\pi\sqrt{3}t^2)$,
which, for our parameters values, is
$x_c \approx 10^{-5}$.
Equivalently, we can say that the SDW is destroyed if
$\mu > \mu_c$,
where the critical chemical potential is
$\mu_c=\Delta^{\text{SDW}}_{\xi0}/2 \approx 0.5$\,meV.
(Allowing for the SDW incommensuration, it is possible to extend the
stability of the SDW, but only slightly.)

\section{Superconductivity mechanism}
%%%%%%%%%%%%%%%%%%%%%%%%%%%%%%%%%%%%%%%%%%%%%%%%%%
\label{BCS}
%%%%%%%%%%%%%%%%%%%%%%%%%%%%%%%%%%%%%%%%%%%%%%%%%%

In this section we analyze properties of the superconducting order that
could arise in doped AA-BLG. Particularly, we will study the regime
$|x|>x_c$,
where the SDW state is completely suppressed.

Our mechanism is based on two theses that are consequences of the structure
of the screened Coulomb interaction. We found in
Sec.~\ref{polarOp}
that the intra-layer RPA repulsion is strong, while the inter-layer
interaction is almost an order of magnitude weaker
$|V^{12}_{\mathbf{p}}|\ll V^{11}_{\mathbf{p}}$.
Moreover,
$V^{12}_{\mathbf{p}}$
can even become negative at small
${\bf p}$.
Thus, we will assume below that (i)~the most stable superconducting order
parameter is one that couples the electrons in different layers. Otherwise,
all Cooper pairs acquire significant positive energy due to high values of
$V_{\bf p}^{ii} > 0$,
see~\ref{intra_layer_appendix}
for extra details.

Additionally, our calculations in
Sec.~\ref{polarOp}
demonstrate that the inter-layer attraction at low transferred momentum is
always weak, and cannot stabilized an experimentally observable
superconducting phase (this is explicitly demonstrated
in~\ref{AppendixA}).
Consequently, (ii)~other sources of attraction must be incorporated into
the model. To satisfy this requirement, we will introduce the
phonon-mediated attraction between the electrons.

\subsection{The phonon-mediated attraction}

We model the BCS-style phonon-mediated attraction by the following
Hamiltonian
\begin{eqnarray}
%%%%%%%%%%%%%%%%%%%%%%%%%%%%%%%%%%%%%%%%%%%%%%%%%%
\label{HamPart}
%%%%%%%%%%%%%%%%%%%%%%%%%%%%%%%%%%%%%%%%%%%%%%%%%%
H_{\text{int}}^{\text{ph}}
&=&
H_{\text{inter}}^{\text{ph}}+H_{\text{intra}}^{\text{ph}},
\\
H_{\text{inter}}^{\text{ph}}
&=&
-\frac{1}{2{\cal N}}\sum_{i\alpha\beta\xi\atop\mathbf{pp}'\sigma \sigma'}
	U_{\mathbf{pp}'}^{\xi\bar{\xi}}
	d^\dag_{\mathbf{K}_{\xi}+\mathbf{p}i\alpha\sigma}
	d^{\phantom{\dag}}_{\mathbf{K}_{\xi}+\mathbf{p}'i\alpha\sigma}
	\times
\nonumber
\\
%%%%%%%%%%%%%%%%%%%%%%%%%%%%%%%%%%%%%%%%%%%%%%%%%%
\label{Hinter}
%%%%%%%%%%%%%%%%%%%%%%%%%%%%%%%%%%%%%%%%%%%%%%%%%%
	&&
	d^\dag_{-{\bf K}_\xi-{\bf p}\bar i\beta \sigma'}
	d^{\phantom{\dag}}_{-{\bf K}_\xi-{\bf p}'\bar i \beta \sigma'},
\\
H_{\text{intra}}^{\text{ph}}
&=&
-\frac{1}{2{\cal N}}\sum_{i\alpha\beta\xi\atop\mathbf{pp}'\sigma \sigma'}
	U_{\mathbf{pp}'}^{\xi\xi}
	d^\dag_{{\bf K}_\xi + {\bf p}i\alpha\sigma}
	d^{\vphantom{\dag}}_{{\bf K}_\xi + {\bf p}'i\alpha\sigma}
	\times
\nonumber
\\
	&&
d^\dag_{{\bf K}_\xi - {\bf p}\bar i \beta \sigma'}
	d^{\vphantom{\dag}}_{{\bf K}_\xi - {\bf p}'\bar i\beta \sigma'},
%%%%%%%%%%%%%%%%%%%%%%%%%%%%%%%%%%%%%%%%%%%%%%%%%%
\label{Hintra}
%%%%%%%%%%%%%%%%%%%%%%%%%%%%%%%%%%%%%%%%%%%%%%%%%%
\end{eqnarray}
where the term
$H_{\text{inter}}^{\text{ph}}$
couples electrons densities in different valleys, while
$H_{\text{intra}}^{\text{ph}}$
couples electrons densities in the same valley. Per our assumption~(i), the
intra-layer interactions are neglected. The inter-layer interaction
parameters are positive
$U_{\mathbf{pp}'}^{\xi\bar\xi}>0$,
$U_{\mathbf{pp}'}^{\xi\xi}>0$,
as they represent the phonon-mediated electron-electron attraction. (The
contribution from the inter-layer screened Coulomb interaction is also
incorporated in these parameters.)

In contrast to AB-BLG where intra-sublattice attraction
dominates~\cite{das_sarma2022supercond_acoustic},
we model our interaction to be sublattice-independent:
$U_{\mathbf{pp}'}^{\xi\bar\xi}$
and
$U_{\mathbf{pp}'}^{\xi\xi}$
lack not only spin but also sublattice indices. Such a dissimilarity
can be explained as follows. For AB-BLG, electrons in low-energy states
avoid the so-called `dimer atoms' of the lattice, localizing as a result on
a single sublattice in each layer. This simplification does not apply when
AA-BLG is studied:
Eq.~(\ref{eigen_En_Fn_1})
demonstrates that all sublattices participate at low energy. Additional
considerations regarding the phonon-mediated attraction can be found in
Sec.~\ref{Discussion}.

To simplify notations, it is convenient to introduce new valley-specific
creation/annihilation band operators
\begin{equation}
%%%%%%%%%%%%%%%%%%%%%%%%%%%%%%%%%%%%%%%%%%%%%%%%%%%
\label{gammaxi}
%%%%%%%%%%%%%%%%%%%%%%%%%%%%%%%%%%%%%%%%%%%%%%%%%%%
\gamma_{\mathbf{p}S\xi\sigma}=\gamma_{\mathbf{K}_{\xi}+\mathbf{p}S\sigma}
\end{equation}
assuming that
$|\mathbf{p}| < K_0$.
Next, we substitute
Eq.~\eqref{eigen_En_Fn_1} into
Eq.~\eqref{HamPart},
and, using the fact that
$\exp(i\varphi_{-\mathbf{K}_{\xi}-\mathbf{p}})
=
\exp(-i\varphi_{\mathbf{K}_{\xi}+\mathbf{p}})$,
$\varphi_{\mathbf{K}_{\xi}+\mathbf{p}}\approx \pi/2+\xi\phi_{\mathbf{p}}$,
and
$\varphi_{\mathbf{K}_{\xi}-\mathbf{p}}
\approx
\pi/2+\xi(\phi_{\mathbf{p}}+\pi)$,
we obtain for
$H_{\text{inter}}^{\text{ph}}$
\begin{widetext}
\begin{eqnarray}
%%%%%%%%%%%%%%%%%%%%%%%%%%%%%%%%%%%%%%%%%%%%%%%%%%
\label{xi_nexi}
%%%%%%%%%%%%%%%%%%%%%%%%%%%%%%%%%%%%%%%%%%%%%%%%%%
H_{\text{inter}}^{\text{ph}}
=
-\frac{1}{16{\cal N}}\sum_{\mathbf{pp}'\xi}\sum_{\sigma\sigma'}
	U_{\mathbf{pp}'}^{\xi\bar{\xi}}
%\times
%\\
%\nonumber
	\left\{
	\left(
		\gamma^{\dag}_{\mathbf{p}2\xi\sigma}
		\!+\!
		\gamma^{\dag}_{\mathbf{p}3\xi\sigma}
	\right)
	\left(
		\gamma^{\phantom{\dag}}_{\mathbf{p}'2\xi\sigma}
		\!+\!
		\gamma^{\phantom{\dag}}_{\mathbf{p}'3\xi\sigma}
	\right)
	\left(
		\gamma^{\dag}_{-\mathbf{p}2\bar{\xi} \sigma'}
		\!-\!
		\gamma^{\dag}_{-\mathbf{p}3\bar{\xi} \sigma'}
	\right)
	\left(
		\gamma^{\phantom{\dag}}_{-\mathbf{p}'2\bar{\xi} \sigma'}
		\!-\!
		\gamma^{\phantom{\dag}}_{-\mathbf{p}'3\bar{\xi} \sigma'}
	\right) +
	\right.
\\
\nonumber
	\left(
		\gamma^{\dag}_{\mathbf{p}2\xi\sigma}
		\!-\!
		\gamma^{\dag}_{\mathbf{p}3\xi\sigma}
	\right)
	\left(
		\gamma^{\phantom{\dag}}_{\mathbf{p}'2\xi\sigma}
		\!-\!
		\gamma^{\phantom{\dag}}_{\mathbf{p}'3\xi\sigma}
	\right)
	\left(
		\gamma^{\dag}_{-\mathbf{p}2\bar{\xi} \sigma'}
		\!+\!
		\gamma^{\dag}_{-\mathbf{p}3\bar{\xi} \sigma'}
	\right)
	\left(
		\gamma^{\phantom{\dag}}_{-\mathbf{p}'2\bar{\xi} \sigma'}
		\!+\!
		\gamma^{\phantom{\dag}}_{-\mathbf{p}'3\bar{\xi} \sigma'}
	\right) +
\\	
\nonumber
 	\left[
	\left(
		\gamma^{\dag}_{\mathbf{p}2\xi\sigma}
		\!+\!
		\gamma^{\dag}_{\mathbf{p}3\xi\sigma}
	\right)
	\left(
		\gamma^{\phantom{\dag}}_{\mathbf{p}'2\xi\sigma}
		\!+\!
		\gamma^{\phantom{\dag}}_{\mathbf{p}'3\xi\sigma}
	\right)
	\left(
		\gamma^{\dag}_{-\mathbf{p}2\bar{\xi} \sigma'}
		\!+\!
		\gamma^{\dag}_{-\mathbf{p}3\bar{\xi} \sigma'}
	\right)
	\left(
		\gamma^{\phantom{\dag}}_{-\mathbf{p}'2\bar{\xi} \sigma'}
		\!+\!
		\gamma^{\phantom{\dag}}_{-\mathbf{p}'3\bar{\xi} \sigma'}
	\right) +
	\right.
\\
\nonumber
	\left.\left.
	\left(
		\gamma^{\dag}_{\mathbf{p}2\xi\sigma}
		\!-\!
		\gamma^{\dag}_{\mathbf{p}3\xi\sigma}
	\right)
	\left(
		\gamma^{\phantom{\dag}}_{\mathbf{p}'2\xi\sigma}
		\!-\!
		\gamma^{\phantom{\dag}}_{\mathbf{p}'3\xi\sigma}
	\right)
	\left(
		\gamma^{\dag}_{-\mathbf{p}2\bar{\xi} \sigma'}
		\!-\!
		\gamma^{\dag}_{-\mathbf{p}3\bar{\xi} \sigma'}
	\right)
	\left(
		\gamma^{\phantom{\dag}}_{-\mathbf{p}'2\bar{\xi} \sigma'}
		\!-\!
		\gamma^{\phantom{\dag}}_{-\mathbf{p}'3\bar{\xi} \sigma'}
	\right)
\right]
%\times
%\right.
%\\
%\nonumber
%\left.
\cos(\phi_{\mathbf{p}}-\phi_{\mathbf{p}'})
\right\}.
\end{eqnarray}
Here, the first two terms describe interaction of electrons on different
layers but identical sublattice. For these terms the phase factors
inherited from
Eq.~(\ref{eigen_En_Fn_1})
cancel each other out. When the electrons inhabit non-identical layers and
non-identical sublattices, their interactions are represented by two other
terms. Since the phases do not cancel for such configurations,
${\cos(\phi_{\mathbf{p}}-\phi_{\mathbf{p}'})}$
emerges. Likewise, for
$H_{\text{intra}}^{\text{ph}}$
we derive
\begin{eqnarray}
%%%%%%%%%%%%%%%%%%%%%%%%%%%%%%%%%%%%%%%%%%%%%%%%%%
\label{xi_xi}
%%%%%%%%%%%%%%%%%%%%%%%%%%%%%%%%%%%%%%%%%%%%%%%%%%
H_{\text{intra}}^{\text{ph}}
=
-\frac{1}{16{\cal N}}\sum_{\mathbf{pp}'\xi} \sum_{\sigma\sigma'}
	U_{\mathbf{pp}'}^{\xi\xi}
%\times
%\\
%\nonumber
	\left\{ \left[
	\left(
		\gamma^{\dag}_{\mathbf{p}2\xi\sigma}
		\!+\!
		\gamma^{\dag}_{\mathbf{p}3\xi\sigma}
	\right)
	\left(
		\gamma^{\phantom{\dag}}_{\mathbf{p}'2\xi\sigma}
		\!+\!
		\gamma^{\phantom{\dag}}_{\mathbf{p}'3\xi\sigma}
	\right)
	\left(
		\gamma^{\dag}_{-\mathbf{p}2\xi \sigma'}
		\!-\!
		\gamma^{\dag}_{-\mathbf{p}3\xi \sigma'}
	\right)
	\left(
		\gamma^{\phantom{\dag}}_{-\mathbf{p}'2\xi \sigma'}
		\!-\!
		\gamma^{\phantom{\dag}}_{-\mathbf{p}'3\xi \sigma'}
	\right) +
	\right.  \right.
\\
\nonumber
	\left.
	\left(
		\gamma^{\dag}_{\mathbf{p}2\xi\sigma}
		\!-\!
		\gamma^{\dag}_{\mathbf{p}3\xi\sigma}
	\right)
	\left(
		\gamma^{\phantom{\dag}}_{\mathbf{p}'2\xi\sigma}
		\!-\!
		\gamma^{\phantom{\dag}}_{\mathbf{p}'3\xi\sigma}
	\right)
	\left(
		\gamma^{\dag}_{-\mathbf{p}2\xi \sigma'}
		\!+\!
		\gamma^{\dag}_{-\mathbf{p}3\xi \sigma'}
	\right)
	\left(
		\gamma^{\phantom{\dag}}_{-\mathbf{p}'2\xi \sigma'}
		\!+\!
		\gamma^{\phantom{\dag}}_{-\mathbf{p}'3\xi \sigma'}
	\right) \right]
%\times
%\\
%\nonumber
	\cos(\phi_{\mathbf{p}}-\phi_{\mathbf{p}'}) +
\\
\nonumber
	\left(
		\gamma^{\dag}_{\mathbf{p}2\xi\sigma}
		\!+\!
		\gamma^{\dag}_{\mathbf{p}3\xi\sigma}
	\right)
	\left(
		\gamma^{\phantom{\dag}}_{\mathbf{p}'2\xi\sigma}
		\!+\!
		\gamma^{\phantom{\dag}}_{\mathbf{p}'3\xi\sigma}
	\right)
	\left(
		\gamma^{\dag}_{-\mathbf{p}2\xi \sigma'}
		\!+\!
		\gamma^{\dag}_{-\mathbf{p}3\xi \sigma'}
	\right)
	\left(
		\gamma^{\phantom{\dag}}_{-\mathbf{p}'2\xi \sigma'}
		\!+\!
		\gamma^{\phantom{\dag}}_{-\mathbf{p}'3\xi \sigma'}
	\right) +
\\
\nonumber
\left.
	\left(
		\gamma^{\dag}_{\mathbf{p}2\xi\sigma}
		\!-\!
		\gamma^{\dag}_{\mathbf{p}3\xi\sigma}
	\right)
	\left(
		\gamma^{\phantom{\dag}}_{\mathbf{p}'2\xi\sigma}
		\!-\!
		\gamma^{\phantom{\dag}}_{\mathbf{p}'3\xi\sigma}
	\right)
	\left(
		\gamma^{\dag}_{-\mathbf{p}2\xi \sigma'}
		\!-\!
		\gamma^{\dag}_{-\mathbf{p}3\xi \sigma'}
	\right)
	\left(
		\gamma^{\phantom{\dag}}_{-\mathbf{p}'2\xi \sigma'}
		\!-\!
		\gamma^{\phantom{\dag}}_{-\mathbf{p}'3\xi \sigma'}
	\right)
\right\},
\end{eqnarray}
\end{widetext}
As in the previous equation, two first terms correspond to different layers
but identical sublattices, while two other terms are for different layers
and different sublattices. Since the phase factors in
Eq.~(\ref{eigen_En_Fn_1})
depend on $\xi$, the phase factor cancellation occurs in the third and
fourth terms.

Note that
Hamiltonians~\eqref{xi_nexi}
and~\eqref{xi_xi}
allow both for singlet and triplet pairing. Below we will limit ourselves
to considering these two symmetries of the order parameter.

\subsection{Mean-field analysis: singlet pairing}
%%%%%%%%%%%%%%%%%%%%%%%%%%%%%%%%%%%%%%%%%%%%%%%%%%
\label{MF_analysis}
%%%%%%%%%%%%%%%%%%%%%%%%%%%%%%%%%%%%%%%%%%%%%%%%%%

Here we investigate the singlet superconducting phase adapting the usual
BCS approach to specific features of AA-BLG. First, we note that the AA-BLG
Fermi surface demonstrates qualitatively different structure for different
doping levels, or, equivalently, for different values of the chemical
potential $\mu$. To avoid studying all possible variations of the problem,
we will limit ourselves to the most physically relevant case of
not-too-strong doping and require that
$|\mu|<t_0$.
Since our AA-BLG model obeys electron-hole symmetry, we consider only the
electron doping,
$\mu>0$.
In this situation, only the bands 2 and 3 cross the Fermi level. Their
Fermi momenta are expressed by
Eqs.~(\ref{kF2})
and~(\ref{kF3}).

We assume that
$U_{\mathbf{pp}'}^{\xi\xi'}$
remain zero unless momenta
$\mathbf{p}$
and
$\mathbf{p}'$
are confined to the rings around the Fermi surfaces of the bands 2 and 3.
The width of these rings is
$q_{\rm D}=\omega_{\rm D}/v_{\rm F}$,
where
$\omega_{\rm D}$
is the Debye frequency. At low doping, when
$k_{\rm F +} - k_{\rm F -} = 2\mu/v_{\rm F} < q_{\rm D}$,
the rings corresponding to different bands intersect. We restrict our
consideration by the case
$\mu > \omega_{\rm D}/2$,
which significantly simplifies calculations. For practical calculations, we
approximate
$U_{\mathbf{pp}'}^{\xi \bar\xi}$
and
$U_{\mathbf{pp}'}^{\xi \xi}$
by constants within these rings
\begin{eqnarray}
%%%%%%%%%%%%%%%%%%%%%%%%%%%%%%%%%%%%%%%%%%%%%%%%%%
\label{FS_rings}
%%%%%%%%%%%%%%%%%%%%%%%%%%%%%%%%%%%%%%%%%%%%%%%%%%
U_{\mathbf{pp}'}^{\xi \bar\xi}
&=&
\left\{
\begin{array}{ll}
	U^{(\nu,\nu')},
	&\;\;|p-k_{F\nu}|<\frac{q_{\rm D}}{2}\wedge|p'-k_{F\nu'}|<\frac{q_{\rm D}}{2}
	\\
	0,&\;\;\text{otherwise}
\end{array}
\right.\!\!\!,
\nonumber
\\
U_{\mathbf{pp}'}^{\xi \xi}
&=&
\left\{
\begin{array}{ll}
	\tilde{U}^{(\nu,\nu')},
	&\;\;|p-k_{F\nu}|<\frac{q_{\rm D}}{2}\wedge|p'-k_{F\nu'}|<\frac{q_{\rm D}}{2}
	\\
	0,&\;\;\text{otherwise}
\end{array}
\right.\!\!\!,
\nonumber
\\
%%%%%%%%%%%%%%%%%%%%%%%%%%%%%%%%%%%%%%%%%%%%%%%%%%
\label{VAnsatz}
%%%%%%%%%%%%%%%%%%%%%%%%%%%%%%%%%%%%%%%%%%%%%%%%%%
\end{eqnarray}
where
$\nu,\,\nu'=\pm1$.
Per
definitions~(\ref{kF2})
and~(\ref{kF3}),
index value
$\nu = +1$
($\nu = -1$)
represents states in
$S=2$
($S=3$)
band. As for the chemical potential, following the above discussion, we
confine $\mu$ to the interval
\begin{eqnarray}
%%%%%%%%%%%%%%%%%%%%%%%%%%%%%%%%%%%%%%%%%%%%%%%%%%
\label{mu_interval}
%%%%%%%%%%%%%%%%%%%%%%%%%%%%%%%%%%%%%%%%%%%%%%%%%%
\frac{\omega_{\rm D}}{2} < \mu < t_0.
\end{eqnarray}
Note that, due to
$\mu_c \sim 1$\,meV
being very small, the SDW is destroyed for all $\mu$ satisfying
inequality~(\ref{mu_interval}).

In general case, in the superconducting state some of the following
expectation values are non-zero:
\begin{equation}
%%%%%%%%%%%%%%%%%%%%%%%%%%%%%%%%%%%%%%%%%%%%%%%%%%%
\label{supercond_eta_gen}
%%%%%%%%%%%%%%%%%%%%%%%%%%%%%%%%%%%%%%%%%%%%%%%%%%%
\eta^{\xi\xi'}_{\mathbf{p}S\sigma\sigma'}=
\langle\gamma^{\phantom{\dag}}_{\mathbf{p}S\xi\sigma}\gamma^{\phantom{\dag}}_{-\mathbf{p}S\xi'\sigma'}\rangle,\;\;S=2,\,3\,.
\end{equation}
For the singlet pairing, considered in this subsection, we should write
\begin{equation}
%%%%%%%%%%%%%%%%%%%%%%%%%%%%%%%%%%%%%%%%%%%%%%%%%%%
\label{supercond_eta_singlet}
%%%%%%%%%%%%%%%%%%%%%%%%%%%%%%%%%%%%%%%%%%%%%%%%%%%
\eta^{\xi\xi'}_{\mathbf{p}S\sigma\sigma'}=\eta^{\xi\xi'}_{\mathbf{p}S}(i\sigma_{y})_{\sigma\sigma'}\,,
\end{equation}
where
\begin{equation}
%%%%%%%%%%%%%%%%%%%%%%%%%%%%%%%%%%%%%%%%%%%%%%%%%%
\label{supercond_eta}
%%%%%%%%%%%%%%%%%%%%%%%%%%%%%%%%%%%%%%%%%%%%%%%%%%
\eta^{\xi\xi'}_{\mathbf{p}S}
=
\langle
	\gamma^{\phantom{\dag}}_{\mathbf{p}S\xi\uparrow}
	\gamma^{\phantom{\dag}}_{-\mathbf{p}S\xi'\downarrow}
\rangle,
\;\;S=2,\,3\,.
\end{equation}
According to
Eq.~\eqref{supercond_eta_singlet},
anomalous matrix element for electrons with the same spin index is zero.

We introduce the order parameters in the form
\begin{equation}
%%%%%%%%%%%%%%%%%%%%%%%%%%%%%%%%%%%%%%%%%%%%%%%%%%
\label{DeltaSC}
%%%%%%%%%%%%%%%%%%%%%%%%%%%%%%%%%%%%%%%%%%%%%%%%%%
\Delta^{\xi\xi'}_{\mathbf{p}S}
=
\frac{1}{4{\cal N}}\sum_{\mathbf{p}'}
	U_{\mathbf{pp}'}^{\xi\xi'}\eta^{\xi\xi'}_{\mathbf{p}'S}\,.
\end{equation}
Let us recall here that the momentum
$\mathbf{p}$
in the electronic operator
$\gamma^{\phantom{\dag}}_{\mathbf{p}S\xi\sigma}$
is counted from the Dirac point
$\mathbf{K}_{\xi}$,
and
$\mathbf{K}_{\bar{\xi}}=-\mathbf{K}_{\xi}$.
The expectation values
$\eta^{\xi\bar{\xi}}_{\mathbf{p}S}$
describe a Cooper pair with zero total momentum, while the total momentum of
pairs represented by expectation values
$\eta^{\xi\xi}_{\mathbf{p}S}$
is equal to
$2\mathbf{K}_{\xi}$.
Thus, the order parameter
$\Delta^{\xi\xi}_{\mathbf{p}S}$
oscillates in real space as
$e^{2i\mathbf{K}_{\xi}\mathbf{r}}$.

To proceed further, we assume that
$\Delta^{\xi\xi'}_{\mathbf{p}S}$
in
Eq.~\eqref{DeltaSC}
depends only on the absolute value of
$\mathbf{p}$.
In this case the terms proportional to
${\cos(\phi_{\mathbf{p}}-\phi_{\mathbf{p}'})}$
in
Eqs.~\eqref{xi_nexi}
and~\eqref{xi_xi}
do not contribute to the mean-field Hamiltonian. Performing the mean-field
decoupling in
Eq.~\eqref{xi_nexi},
we obtain the mean-field form of
$H^{\rm ph}_{\rm inter}$
\begin{equation}
%%%%%%%%%%%%%%%%%%%%%%%%%%%%%%%%%%%%%%%%%%%%%%%%%%
\label{xi_nexi_MF}
%%%%%%%%%%%%%%%%%%%%%%%%%%%%%%%%%%%%%%%%%%%%%%%%%%
H_{\text{inter}}^{\text{MF}}
=
\sum_{\mathbf{p}\xi}
	\Delta_{{\bf p} \xi}\!
	\left(
		\gamma^{\dag}_{\mathbf{p}2\xi\uparrow}
		\gamma^{\dag}_{-\mathbf{p}2\bar{\xi}\downarrow}
		-
		\gamma^{\dag}_{\mathbf{p}3\xi\uparrow}
		\gamma^{\dag}_{-\mathbf{p}3\bar{\xi}\downarrow}
	\right)
%\\
%\nonumber
+ {\rm H.c.} + C,
\end{equation}
where the $c$-number constant is
\begin{eqnarray}
C
=
\frac{1}{4{\cal N}}\sum_{\mathbf{pp}'\xi}
	U_{\mathbf{pp}'}^{\xi\bar{\xi}}
	\left(
		\eta^{\xi\bar{\xi}*}_{\mathbf{p}2}
		-
		\eta^{\xi\bar{\xi}*}_{\mathbf{p}3}
	\right)
	\left(
		\eta^{\xi\bar{\xi}}_{\mathbf{p}'2}
		-
		\eta^{\xi\bar{\xi}}_{\mathbf{p}'3}
	\right),
\end{eqnarray}
and the inter-layer order parameter is
$\Delta_{{\bf p} \xi}
=
\Delta^{\xi\bar{\xi}}_{\mathbf{p}2}
-
\Delta^{\xi\bar{\xi}}_{\mathbf{p}3}$.
Of course, intra-layer order parameter can be defined as well. However,
strong intra-layer repulsion disfavors such order,
see~\ref{intra_layer_appendix}
for more details.

Similar to derivations above, introducing
$\tilde{\Delta}_{\mathbf{p}\xi}
=
\Delta^{\xi\xi}_{\mathbf{p}2}+\Delta^{\xi\xi}_{\mathbf{p}3}$
and
\begin{eqnarray}
\Tilde C
=
\frac{1}{4{\cal N}}\sum_{\mathbf{pp}'\xi}
	U_{\mathbf{pp}'}^{\xi\xi}
	\left(
		\eta^{\xi\xi*}_{\mathbf{p}2}
		+
		\eta^{\xi\xi*}_{\mathbf{p}3}
	\right)
	\left(
		\eta^{\xi\xi}_{\mathbf{p}'2}
		+
		\eta^{\xi\xi}_{\mathbf{p}'3}
	\right),
\end{eqnarray}
we express the mean-field form of
$H_{\text{intra}}^{\rm ph}$
as follows
\begin{equation}
%%%%%%%%%%%%%%%%%%%%%%%%%%%%%%%%%%%%%%%%%%%%%%%%%%
\label{xi_xi_MF}
%%%%%%%%%%%%%%%%%%%%%%%%%%%%%%%%%%%%%%%%%%%%%%%%%%
H_{\text{intra}}^{\text{MF}}
=
\sum_{\mathbf{p}\xi}
	\tilde{\Delta}_{\mathbf{p}\xi}
	\left(
		\gamma^{\dag}_{\mathbf{p}2\xi\uparrow}
		\gamma^{\dag}_{-\mathbf{p}2\xi\downarrow}
		\!+\!
		\gamma^{\dag}_{\mathbf{p}3\xi\uparrow}
		\gamma^{\dag}_{-\mathbf{p}3\xi\downarrow}
	\right)
%\\
%\nonumber
+ {\rm H.c.} + \Tilde C.
\end{equation}
%\end{widetext}
It is convenient to define the eight-component operator-valued spinor
\begin{eqnarray}
%%%%%%%%%%%%%%%%%%%%%%%%%%%%%%%%%%%%%%%%%%%%%%%%%%
\label{Psi}
%%%%%%%%%%%%%%%%%%%%%%%%%%%%%%%%%%%%%%%%%%%%%%%%%%
\Psi^{\dag}_{\mathbf{p}}
&=&
\left(
	\Psi^{\dag}_{{\bf p} +1 \uparrow},\,
	\Psi^{\dag}_{{\bf p} -1 \uparrow},\,
	\Psi^{\vphantom{\dag}}_{-{\bf p} +1 \downarrow},\,
	\Psi^{\vphantom{\dag}}_{-{\bf p} -1 \downarrow}
\right),
\end{eqnarray}
where the two-component spinor
$\Psi^{\vphantom{\dag}}_{{\bf p} \xi \sigma}$
equals
\begin{eqnarray}
\Psi^{\vphantom{\dag}}_{{\bf p} \xi \sigma}
=
\left(
	\gamma^{\phantom{\dag}}_{\mathbf{p}2 \xi \sigma},\,
	\gamma^{\phantom{\dag}}_{\mathbf{p}3 \xi \sigma}
\right).
\end{eqnarray}
Using
$\Psi^{\dag}_{\mathbf{p}}$
and
$\Psi^{\vphantom{\dagger}}_{\mathbf{p}}$,
we write the total mean-field Hamiltonian in the form
\begin{eqnarray}
%%%%%%%%%%%%%%%%%%%%%%%%%%%%%%%%%%%%%%%%%%%%%%%%%%
\label{HMF}
%%%%%%%%%%%%%%%%%%%%%%%%%%%%%%%%%%%%%%%%%%%%%%%%%%
H^{\text{MF}}
=
\sum_{\mathbf{p}}
	\Psi^{\dag}_{\mathbf{p}}
	\hat{H}^{\vphantom{\dagger}}_{\mathbf{p}}
	\Psi^{\vphantom{\dag}}_{\mathbf{p}}+
C + \Tilde C,
\end{eqnarray}
where the
$8\times8$
matrix
$\hat{H}_{\mathbf{p}}$
reads
\begin{widetext}
\begin{equation}
%%%%%%%%%%%%%%%%%%%%%%%%%%%%%%%%%%%%%%%%%%%%%%%%%%
\label{hatH}
%%%%%%%%%%%%%%%%%%%%%%%%%%%%%%%%%%%%%%%%%%%%%%%%%%
\hat{H}_{\mathbf{p}}
=
\left(\begin{array}{cccccccc}
\varepsilon_{\mathbf{p}}-\mu&0&0&0&\tilde{\Delta}_{\mathbf{p}+1}&0&\Delta_{\mathbf{p}+1}&0\\
0&-\varepsilon_{\mathbf{p}}-\mu&0&0&0&\tilde{\Delta}_{\mathbf{p}+1}&0&-\Delta_{\mathbf{p}+1}\\
0&0&\varepsilon_{\mathbf{p}}-\mu&0&\Delta_{\mathbf{p}-1}&0&\tilde{\Delta}_{\mathbf{p}-1}&0\\
0&0&0&-\varepsilon_{\mathbf{p}}-\mu&0&-\Delta_{\mathbf{p}-1}&0&\tilde{\Delta}_{\mathbf{p}-1}\\
\tilde{\Delta}^{*}_{\mathbf{p}+1}&0&\Delta^{*}_{\mathbf{p}-1}&0&-\varepsilon_{\mathbf{p}}+\mu&0&0&0\\
0&\tilde{\Delta}^{*}_{\mathbf{p}+1}&0&-\Delta^{*}_{\mathbf{p}-1}&0&\varepsilon_{\mathbf{p}}+\mu&0&0\\
\Delta^{*}_{\mathbf{p}+1}&0&\tilde{\Delta}^{*}_{\mathbf{p}-1}&0&0&0&-\varepsilon_{\mathbf{p}}+\mu&0\\
0&-\Delta^{*}_{\mathbf{p}+1}&0&\tilde{\Delta}^{*}_{\mathbf{p}-1}&0&0&0&\varepsilon_{\mathbf{p}}+\mu
\end{array}\right),
\end{equation}
\end{widetext}
This matrix depends on four order parameters
$\Delta_{\mathbf{p}\xi}$
and
$\tilde{\Delta}_{\mathbf{p}\xi}$
with
$\xi=\pm1$.
However, not all of theses quantities are independent. Indeed, in the
singlet state, according to the
definition~\eqref{supercond_eta_singlet},
the following conditions hold true for the inter-valley and intra-valley order parameters
\begin{eqnarray}
%%%%%%%%%%%%%%%%%%%%%%%%%%%%%%%%%%%%%%%%%%%%%%%%%%
\label{symm_Delta_singlet}
%%%%%%%%%%%%%%%%%%%%%%%%%%%%%%%%%%%%%%%%%%%%%%%%%%
\Delta_{\mathbf{p}\xi}=\Delta_{-\mathbf{p}\bar{\xi}}\,,
\\
%%%%%%%%%%%%%%%%%%%%%%%%%%%%%%%%%%%%%%%%%%%%%%%%%%
\label{symm_tildeDelta_singlet}
%%%%%%%%%%%%%%%%%%%%%%%%%%%%%%%%%%%%%%%%%%%%%%%%%%
\tilde{\Delta}_{\mathbf{p}\xi}=\tilde{\Delta}_{-\mathbf{p}\xi}\,.
\end{eqnarray}
The first of these equations is a consequence of the relation
$\mathbf{K}_{\bar{\xi}}=-\mathbf{K}_{\xi}$.
Since we assumed above that
$\Delta_{\mathbf{p}\xi}$
does not depend on the direction of
$\mathbf{p}$,
one can simplify
\begin{equation}
%%%%%%%%%%%%%%%%%%%%%%%%%%%%%%%%%%%%%%%%%%%%%%%%%%
\label{symm_Delta_singlet_s}
%%%%%%%%%%%%%%%%%%%%%%%%%%%%%%%%%%%%%%%%%%%%%%%%%%
\Delta_{\mathbf{p}\xi}
=
\Delta_{\mathbf{p}\bar{\xi}}\equiv\Delta_{\mathbf{p}}\,.
\end{equation}
As for intra-valley pairing, no symmetry condition similar to
Eq.~\eqref{symm_Delta_singlet}
can be written for
$\tilde{\Delta}_{\mathbf{p}\xi}$
and
$\tilde{\Delta}_{\mathbf{p}\bar{\xi}}$.
However, since both electrons in a Cooper pair have identical valley index
values, we assume the following relation for the intra-valley order
parameters
\begin{equation}
%%%%%%%%%%%%%%%%%%%%%%%%%%%%%%%%%%%%%%%%%%%%%%%%%% %
\label{symm_tildeDelta_singlet_s}
%%%%%%%%%%%%%%%%%%%%%%%%%%%%%%%%%%%%%%%%%%%%%%%%%% %
\tilde{\Delta}_{\mathbf{p}+1}=e^{i\chi_{\mathbf{p}}}\tilde{\Delta}_{\mathbf{p}-1}\equiv\tilde{\Delta}_{\mathbf{p}}\,,
\end{equation}
where the phase shift $\chi_{\mathbf{p}}$ should be found self-consistently together with $\Delta_{\mathbf{p}}$ and $\tilde{\Delta}_{\mathbf{p}}$.

The quasiparticle spectrum is obtained by the diagonalization of the
matrix~\eqref{hatH}.
Assuming that
$\Delta_{\mathbf{p}}$
and
$\tilde{\Delta}_{\mathbf{p}}$
are the real-valued functions of
$\mathbf{p}$,
we obtain
\begin{equation}
%%%%%%%%%%%%%%%%%%%%%%%%%%%%%%%%%%%%%%%%%%%%%%%%%% %
\label{EcosChi}
%%%%%%%%%%%%%%%%%%%%%%%%%%%%%%%%%%%%%%%%%%%%%%%%%% %
E^{(1,2,...,8)}_{\mathbf{p}}=
\mp\sqrt{\left(\varepsilon_{\mathbf{p}}\mp\mu\right)^2+
\Delta^2_{\mathbf{p}}+\tilde{\Delta}^2_{\mathbf{p}}\pm
2\Delta_{\mathbf{p}}\tilde{\Delta}_{\mathbf{p}}\cos\frac{\chi_{\mathbf{p}}}{2}}.
\end{equation}
Minimizing the grand potential corresponding to the
Hamiltonian~\eqref{HMF}
one derives the self-consistency equations for the superconducting
order parameters. Since
$\eta^{\xi\bar{\xi}}_{\mathbf{p}S}$
appears in Hamiltonian~\eqref{HMF}
only in combination
$\eta^{\xi\bar{\xi}}_{\mathbf{p}2}-\eta^{\xi\bar{\xi}}_{\mathbf{p}3}$,
and
$\eta^{\xi\xi}_{\mathbf{p}S}$
appears in combination
$\eta^{\xi\xi}_{\mathbf{p}2}+\eta^{\xi\xi}_{\mathbf{p}3}$,
then, only
$\Delta_{\mathbf{p}\xi}$
and
$\Tilde \Delta_{\mathbf{p}\xi}$
enter the self-consistency equations. Taking into account the
relationships~\eqref{symm_Delta_singlet_s},
\eqref{symm_tildeDelta_singlet_s},
and~\eqref{EcosChi},
the system of self-consistency equations can be expressed as
\begin{eqnarray}
%%%%%%%%%%%%%%%%%%%%%%%%%%%%%%%%%%%%%%%%%%%%%%%%%%
\label{DeltaEq}
%%%%%%%%%%%%%%%%%%%%%%%%%%%%%%%%%%%%%%%%%%%%%%%%%%
\Delta_{\mathbf{p}}
\!&=&\!
\frac{1}{16}\!\sum_{\tau\zeta}\!\!
	\int\frac{d^2\mathbf{p}'}{v_{\rm BZ}}
		\frac{U_{\mathbf{pp}'}^{\xi \bar\xi}
			(\Delta_{\mathbf{p}'}
			+
			\zeta\tilde{\Delta}_{\mathbf{p}'}
			\cos\frac{\chi_{\mathbf{p}'}}{2}
			)
		}
		{\sqrt{(\varepsilon_{\mathbf{p}'}-\tau\mu)^2
			+
			\big[\Delta^{(\zeta)}_{\mathbf{p}'}\big]^2
			}
		}
\!\times
\nonumber
\\
&&
\tanh\left(
	\frac{\sqrt{(\varepsilon_{\mathbf{p}'}-\tau\mu)^2+
\big[\Delta^{(\zeta)}_{\mathbf{p}'}\big]^2}}{2T}
\right),
\\
%%%%%%%%%%%%%%%%%%%%%%%%%%%%%%%%%%%%%%%%%%%%%%%%%%
\label{tildeDeltaEq}
%%%%%%%%%%%%%%%%%%%%%%%%%%%%%%%%%%%%%%%%%%%%%%%%%%
\tilde{\Delta}_{\mathbf{p}}
\!&=&\!
\frac{1}{16}\!
\sum_{\tau\zeta}\!\!
	\int\frac{d^2\mathbf{p}'}{v_{\rm BZ}}
		\frac{U_{\mathbf{pp}'}^{\xi \xi}
			(\tilde{\Delta}_{\mathbf{p}'}
			+
			\zeta\Delta_{\mathbf{p}'}
			\cos\frac{\chi_{\mathbf{p}'}}{2}
			)
		}
	 {\sqrt{(\varepsilon_{\mathbf{p}'}-\tau\mu)^2
	+
	\big[\Delta^{(\zeta)}_{\mathbf{p}'}\big]^2
	}}
\!\times
\nonumber
\\
&&
\tanh\left(
	 \frac{\sqrt{(\varepsilon_{\mathbf{p}'}-\tau\mu)^2+
\big[\Delta^{(\zeta)}_{\mathbf{p}'}\big]^2}}{2T}
\right),
\end{eqnarray}
where
$\big[\Delta^{(\zeta)}_{\mathbf{p}}\big]^2
=
\Delta^2_{\mathbf{p}}+\tilde{\Delta}^2_{\mathbf{p}}
+
2\zeta\Delta_{\mathbf{p}}\tilde{\Delta}_{\mathbf{p}}
\cos\frac{\chi_{\mathbf{p}}}{2}$,
the integration is executed over the rings in which the interactions are
finite, as defined by
Eq.~(\ref{FS_rings}),
and the summations are performed over binary indices
$\tau, \zeta = \pm 1$.
In principle, we also should write the self-consistently equation for the
phase shift
$\chi_{\mathbf{p}}$.
However, as we will see below, it is not necessary.

The analysis of
Eqs.~\eqref{DeltaEq}
and~\eqref{tildeDeltaEq}
shows that
$\Delta_{\mathbf{p}}\neq0$,
$\tilde{\Delta}_{\mathbf{p}}=0$
and
$\Delta_{\mathbf{p}}=0$,
$\tilde{\Delta}_{\mathbf{p}}\neq0$
are the solutions to the
equations~\eqref{DeltaEq}
and~\eqref{tildeDeltaEq}
for arbitrary
$\chi_{\mathbf{p}}$.
Both these solutions correspond to local energy minima, one is stable,
another one is metastable. If the intra-valley coupling
$U_{\mathbf{pp}'}^{\xi \bar\xi}$
dominates, then the solution
$\Delta_{\mathbf{p}}\neq0$,
$\tilde{\Delta}_{\mathbf{p}}=0$
represent the stable minimum, with the lowest energy. Otherwise, we have to
choose the solution
$\Delta_{\mathbf{p}}=0$,
$\tilde{\Delta}_{\mathbf{p}}\neq0$.
In the latter case, the phase shift can be arbitrary, and the ground state
energy is independent of
$\chi_{\mathbf{p}}$.
We did not find any solutions (corresponding to the energy minimum) when
both
$\Delta_{\mathbf{p}}$
and
$\tilde{\Delta}_{\mathbf{p}}$
are non-zero. Thus, the order parameters
$\Delta_{\mathbf{p}}$
and
$\tilde{\Delta}_{\mathbf{p}}$
compete with each other.

When the coupling constants
%$U_{\mathbf{pp}'}$ and $\tilde{U}_{\mathbf{pp}'}$
satisfy
Eq.~(\ref{VAnsatz}),
the order parameters
$\Delta_{\mathbf{p}}$
and
$\tilde{\Delta}_{\mathbf{p}}$
have the following structure
\begin{eqnarray}
\Delta_{\mathbf{p}}
&=&
\left\{
\begin{array}{ll}
	\Delta^{(\nu)},
	&\;\;|p-k_{F\nu}|<\frac{q_{\rm D}}{2}
	\\
	0,&\;\;\text{otherwise}
\end{array}
\right.\!\!\!,
\nonumber
\\
\tilde{\Delta}_{\mathbf{p}}
&=&
\left\{
\begin{array}{ll}
	\tilde{\Delta}^{(\nu)}, &\;\;|p-k_{F\nu}|<\frac{q_{\rm D}}{2}\\
	0,&\;\;\text{otherwise}
\end{array}
\right.\!\!\!.
%%%%%%%%%%%%%%%%%%%%%%%%%%%%%%%%%%%%%%%%%%%%%%%%%%
\label{DeltaAnsatz}
%%%%%%%%%%%%%%%%%%%%%%%%%%%%%%%%%%%%%%%%%%%%%%%%%%
\end{eqnarray}
Thus, we consider $s$-wave pairing. This choice is dictated by the simple
BSC-like form of the interaction functions
$U_{\mathbf{pp}'}^{\xi\xi'}$,
Eq.~\eqref{VAnsatz}.
In principle other types of pairing can win if we consider interaction
$U_{\mathbf{pp}'}^{\xi\xi'}$
with more complicated structure.

Substituting
Eqs.~\eqref{VAnsatz}
and~\eqref{DeltaAnsatz}
into
Eqs.~\eqref{DeltaEq}
and~\eqref{tildeDeltaEq}
one obtains the self-consistency equations for
$\Delta^{(\nu)}$
and
$\tilde{\Delta}^{(\nu)}$.
Linearized forms of the self-consistency equations can be used to calculate
the superconducting critical temperature
$T_{\rm sc}$.
Recall that, when $T$ is close to
$T_{\rm sc}$,
order parameters
$\Delta_{\mathbf{p}}$
and
$\tilde{\Delta}_{\mathbf{p}}$
under the square roots in
Eqs.~\eqref{DeltaEq}
and~\eqref{tildeDeltaEq}
can be set equal to zero. The remaining expression becomes a linear
integral equation for
$\Delta_{\mathbf{p}}$
and
$\tilde{\Delta}_{\mathbf{p}}$.

Substituting
Eq.~\eqref{DeltaAnsatz}
into
Eqs.~\eqref{DeltaEq}
and~\eqref{tildeDeltaEq},
performing the summation over $\tau$ and $\zeta$, and taking the
integrals over
$\mathbf{p}$
in the limit
$T\ll\omega_{\rm D}$,
one transforms the latter integral equation into a homogeneous system of
linear equations for
$\Delta^{(\nu)}$
and
$\tilde{\Delta}^{(\nu)}$
\begin{eqnarray}
%%%%%%%%%%%%%%%%%%%%%%%%%%%%%%%%%%%%%%%%%%%%%%%%%%
\label{DeltaLin0}
%%%%%%%%%%%%%%%%%%%%%%%%%%%%%%%%%%%%%%%%%%%%%%%%%%
\Delta^{(\nu)}
&=&
\sum_{\nu'}
	\lambda^{(\nu,\nu')}\Delta^{(\nu')}
	\ln\frac{\omega_{\rm D}e^{\omega_{\rm D}/4\mu}}{\varkappa T},
\\
%%%%%%%%%%%%%%%%%%%%%%%%%%%%%%%%%%%%%%%%%%%%%%%%%%
\label{DeltaLin}
%%%%%%%%%%%%%%%%%%%%%%%%%%%%%%%%%%%%%%%%%%%%%%%%%%
\tilde{\Delta}^{(\nu)}
&=&
\sum_{\nu'}
	\tilde{\lambda}^{(\nu,\nu')} \tilde{\Delta}^{(\nu')}
	\ln\frac{\omega_{\rm D}e^{\omega_{\rm D}/4\mu}}{\varkappa T},
\end{eqnarray}
where the dimensionless coupling constants are
\begin{equation}
%%%%%%%%%%%%%%%%%%%%%%%%%%%%%%%%%%%%%%%%%%%%%%%%%%
\label{lambda_defs}
%%%%%%%%%%%%%%%%%%%%%%%%%%%%%%%%%%%%%%%%%%%%%%%%%%
\lambda^{(\nu,\nu')}=\frac{\pi k_{F\nu'}}{2v_{\rm BZ}v_{\rm F}}U^{(\nu,\nu')},
\;\;
\tilde{\lambda}^{(\nu,\nu')}
=
\frac{\pi k_{F\nu'}}{2v_{\rm BZ}v_{\rm F}}\tilde{U}^{(\nu,\nu')}.
\end{equation}
and
$\varkappa=\pi e^{-C}\approx 1.764$
($C$
is the Euler's constant). The factor
$e^{\omega_{\rm D}/4\mu}$
under logarithms in
Eqs.~\eqref{DeltaLin0}
and~\eqref{DeltaLin}
arises due to the terms with
$\tau=-\nu'$.

Systems of
equations~(\ref{DeltaLin0})
and~(\ref{DeltaLin})
for
$\Delta^{(\nu)}$
and
$\tilde{\Delta}^{(\nu)}$
are decoupled from each other. Note that for generic $T$ both systems have
trivial solution
$\Delta^{(\nu)} = 0$
and
$\tilde{\Delta}^{(\nu)} = 0$
only. There are, however, four values of $T$ [two values
for~(\ref{DeltaLin0})
and two more
for~(\ref{DeltaLin})]
for which a non-trivial solution is possible. The transition temperature
corresponds to the largest of these four.

\begin{figure}
\centering
\includegraphics[width=0.4\columnwidth]{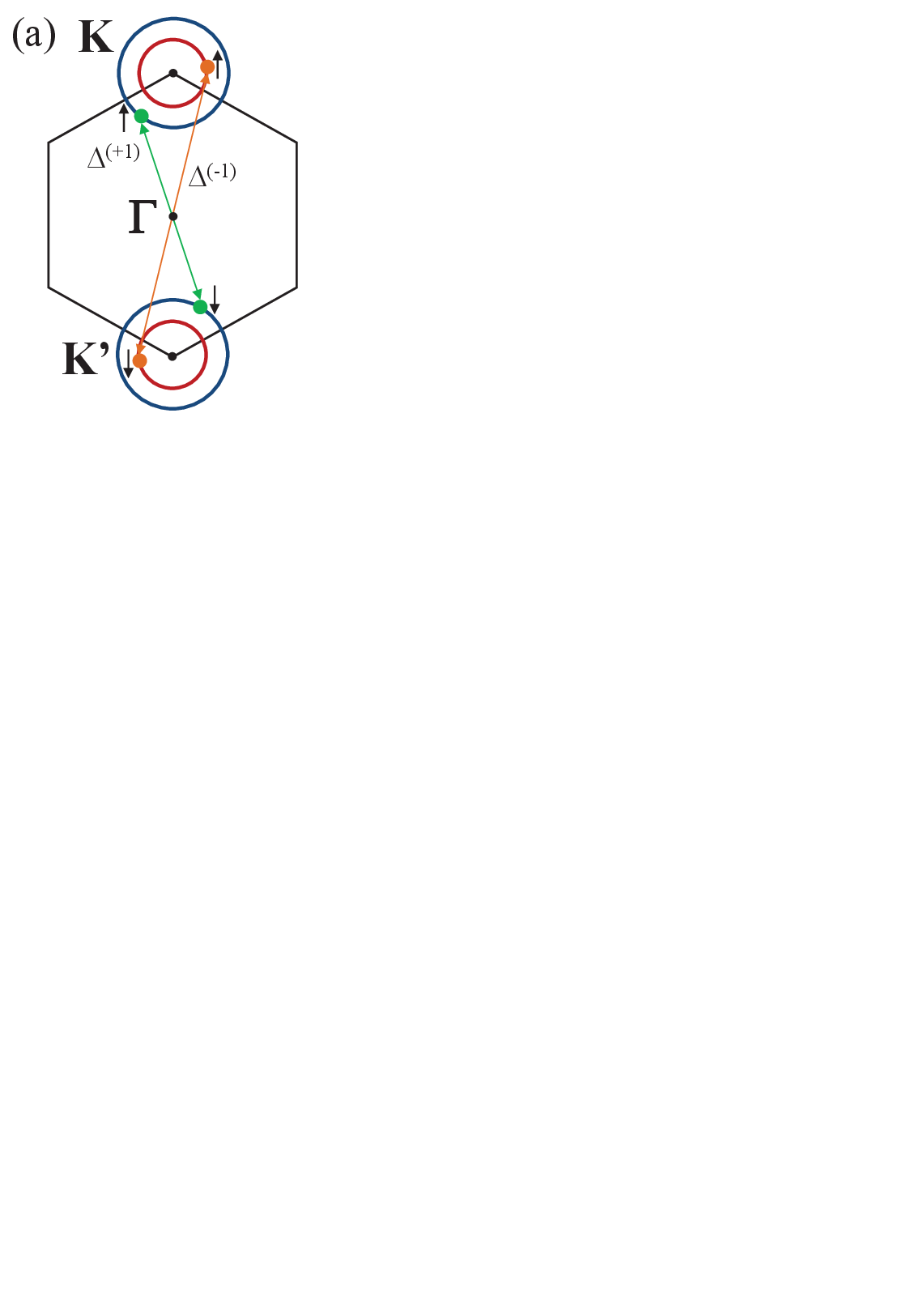}
\qquad
\includegraphics[width=0.4\columnwidth]{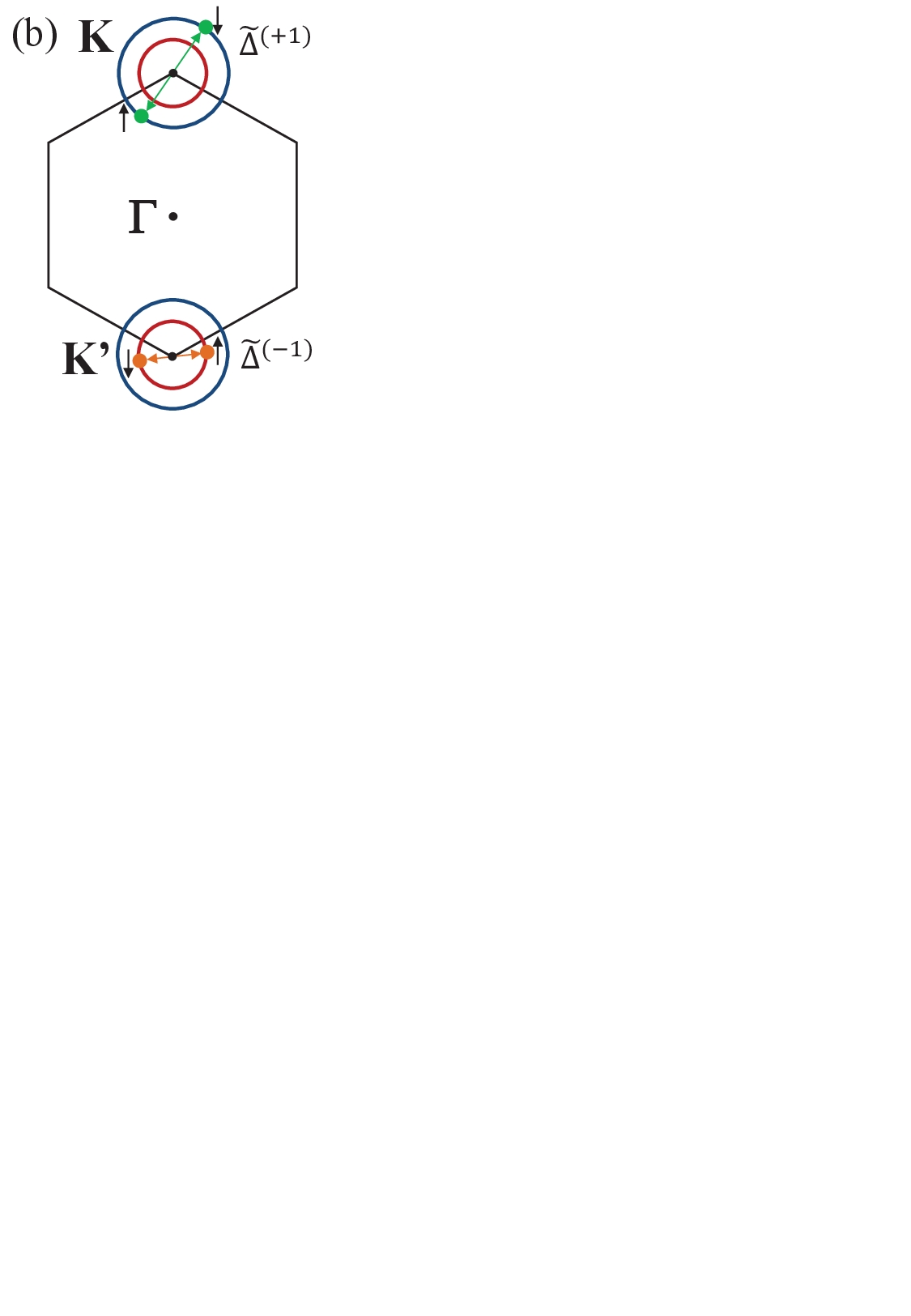}
\caption{A schematic illustration of the possible types of pairing
considered in the paper. (a)~Pairing of the electrons from different
valleys. Corresponding coupling constant is
$U^{\xi \bar\xi}_{{\bf p} {\bf p}'}$,
total Cooper pair momentum is zero.
(b)~Pairing of the electrons within a single valley. The superconductivity
phase is controlled by coupling
$U^{\xi \xi}_{{\bf p} {\bf p}'}$.
In such a situation, the momentum of a Cooper pair is
$\pm 2{\bf K}_1$,
depending on valley index. On both panels hexagons
represent graphene Brillouin zone. Larger (blue) circles correspond to
electronic Fermi surface sheets, smaller (red) circles correspond to hole
sheets. (Fermi momenta drawn not to scale.) Possible Cooper pairs are
marked by (red and green) double-arrow lines.
%%%%%%%%%%%%%%%%%%%%%%%%%%%%%%%%%%%%%%%%%%%%%%%%%%
\label{Fig_pairing}
%%%%%%%%%%%%%%%%%%%%%%%%%%%%%%%%%%%%%%%%%%%%%%%%%%
}
\end{figure}

To be more specific, let us consider the case
$\Delta^{(\nu)} \ne 0$,
$\tilde{\Delta}^{(\nu)}=0$.
This represents the ordering with zero total momentum, see panel~(a) of
Fig.~\ref{Fig_pairing}.
When
$U^{(\nu,\bar{\nu})}=0$,
that is, there is no coupling between order parameters corresponding to
different bands, there are two candidates to become the transition
temperature
\begin{eqnarray}
T^{(\nu)} = \varkappa^{-1} \tilde \omega_{\rm D}e^{-1/\lambda^{(\nu,\nu)}},
\quad
\nu = \pm 1.
\end{eqnarray}
Here the energy scale
$\tilde \omega_{\rm D} = \omega_{\rm D} e^{\omega_{\rm D}/4\mu}$
remains of the order of
$\omega_{\rm D}$,
as long as
condition~(\ref{mu_interval})
is satisfied.

As temperature grows,
$\Delta^{(+1)}$
and
$\Delta^{(-1)}$
vanish independently. First vanishes the order parameter with the smallest
$\lambda^{(\nu,\nu)}$.
When
$U^{(\nu,\nu)}=U_0$,
we have
$T^{(-1)}<T^{(+1)}$,
thus
$\Delta^{(-1)}$
is the first to nullify. This is because
$\nu = -1$
represents the hole band that has smaller density of states. If we assume
that
$U^{(\nu,\nu)}$
are doping independent, then
$T^{(+1)}$
increases when $\mu$ grows, while
$T^{(-1)}$
decreases down to zero at
$\mu=t_0$.

When
$U^{(\nu,\bar{\nu})}\neq0$,
there is a single transition temperature, and the order parameters
$\Delta^{(+1)}$
and
$\Delta^{(-1)}$
vanish simultaneously. This happens at
$T_{\rm sc} = \varkappa^{-1} \tilde \omega_{\rm D} e^{-1/\Lambda}$,
where
\begin{eqnarray}
\Lambda
&=&
\frac12\left[
	\lambda^{(1,1)}+\lambda^{(-1,-1)}
	+\phantom{\sqrt{(\lambda^{(1,1)}})^2}
\right.
\nonumber
\\
&&\left.
	\sqrt{
		[\lambda^{(1,1)}-\lambda^{(-1,-1)}]^2
		+
		4\lambda^{(1,-1)}\lambda^{(-1,1)}
	}
\right].
\end{eqnarray}
Let us now analyze the behavior of $\Lambda$ with doping. For simplicity we
consider only the case when
$U^{(\nu,\nu)}=U_0$
and
$U^{(\nu,\bar{\nu})}=U_1$,
where
$U_{0,1}$
are independent of doping. In this case we have
\begin{equation}
\Lambda=\lambda_0+\sqrt{\lambda_1^2+\mu^2(\lambda_0^2-\lambda_1^2)/t_0^2}\,,
\end{equation}
where
$\lambda_{0,1}=\pi t_0U_{0,1}/(2v_{\rm BZ}v_{\rm F}^2)$.
If
$\lambda_0>\lambda_1$,
that is, the intra-band interaction is greater than inter-band one, the
coupling $\Lambda$ (and the transition temperature) increases when doping
grows. When
$\lambda_0<\lambda_1$,
the coupling $\Lambda$ decreases when doping increases. Finally, at
$\lambda_0=\lambda_1$,
the $\Lambda$ is doping independent, and the transition temperature only
slightly decreases with doping due to the exponent
$e^{\omega_{\rm D}/4\mu}$.

The case
$\Tilde \Delta_{\mathbf{p}}\neq0$
and
$\Delta_{\mathbf{p}}=0$
is schematically represented in panel~(b) of
Fig.~\ref{Fig_pairing}.
Such a case is of particular interest as it corresponds to the Cooper pairs
with non-zero total momentum. As long as we ignore trigonal warping, the
analysis presented in this section for
$\Delta_{\bf p}$
can be applied with little modifications to
$\Tilde \Delta_{\bf p}$
as well. On the other hand, trigonal warping breaks perfect inversion
symmetry of a circular Fermi surface sheet, making single-electron states
$\gamma^{\phantom{\dag}}_{\mathbf{p}S \xi \uparrow}$
and
$\gamma^{\phantom{\dag}}_{\mathbf{-p}S \xi \downarrow}$
non-degenerate. This leads to suppression of this type of
superconductivity. Such an issue does not affect
$\Delta_{\bf p}$
as the time-reversal symmetry guarantees that
$\gamma^{\phantom{\dag}}_{\mathbf{p}S \xi \uparrow}$
and
$\gamma^{\phantom{\dag}}_{\mathbf{-p}S \bar\xi \downarrow}$
remain degenerate. Thus, the trigonal warping is a factor favoring
$\Delta_{\bf p}$
over
$\Tilde \Delta_{\bf p}$.

\subsection{Triplet pairing}

For triplet pairing the expectation values
$\eta^{\xi\xi'}_{\mathbf{p}S\sigma\sigma'}$
[see
Eq.~\eqref{supercond_eta_gen}]
can be written in the form
\begin{equation}
%%%%%%%%%%%%%%%%%%%%%%%%%%%%%%%%%%%%%%%%%%%%%%%%%%
\label{supercond_eta_triplet}
%%%%%%%%%%%%%%%%%%%%%%%%%%%%%%%%%%%%%%%%%%%%%%%%%%
\eta^{\xi\xi'}_{\mathbf{p}S\sigma\sigma'}
=
(i\bm{\eta}^{\xi\xi'}_{\mathbf{p}S} \bm{\sigma}\sigma_{y})_{\sigma\sigma'}
\,,
\end{equation}
where instead of $\eta^{\xi\xi'}_{\mathbf{p}S}$,
Eq.~\eqref{supercond_eta},
we introduce vector-valued functions
$\bm{\eta}^{\xi\xi'}_{\mathbf{p}S}$.
The order parameter
\begin{equation}
%%%%%%%%%%%%%%%%%%%%%%%%%%%%%%%%%%%%%%%%%%%%%%%%%%
\label{DeltaSCtriplet}
%%%%%%%%%%%%%%%%%%%%%%%%%%%%%%%%%%%%%%%%%%%%%%%%%%
\bm{\Delta}^{\xi\xi'}_{\mathbf{p}S}=
\frac{1}{4{\cal N}}\sum_{\mathbf{p}'}
	U_{\mathbf{pp}'}^{\xi\xi'}\bm{\eta}^{\xi\xi'}_{\mathbf{p}'S}
\end{equation}
is a vector as well.
%Since all coefficients in self-consistency equations  derived below are purely real, ${\Delta}^{\xi\xi'}_{\mathbf{p}S}$ can be chosen to be real.

In the triplet state, the inter-valley order parameters
$\bm{\Delta}_{\mathbf{p}\xi}
=
\bm{\Delta}^{\xi\bar{\xi}}_{\mathbf{p}2}
-
\bm{\Delta}^{\xi\bar{\xi}}_{\mathbf{p}3}$
should satisfy the relation
\begin{equation}
%%%%%%%%%%%%%%%%%%%%%%%%%%%%%%%%%%%%%%%%%%%%%%%%%%
\label{symm_Delta_triplet}
%%%%%%%%%%%%%%%%%%%%%%%%%%%%%%%%%%%%%%%%%%%%%%%%%%
\bm{\Delta}_{\mathbf{p}\xi}=-\bm{\Delta}_{-\mathbf{p}\bar{\xi}}\,.
\end{equation}
As before, one can show that for the functions
$U_{\mathbf{pp}'}^{\xi\xi'}$
in the form of
Eq.~\eqref{FS_rings},
the minimum of energy corresponds to the case when
$\bm{\Delta}_{\mathbf{p}\xi}$
is insensitive to the direction of
$\mathbf{p}$.
In this situation one can write
\begin{equation}
%%%%%%%%%%%%%%%%%%%%%%%%%%%%%%%%%%%%%%%%%%%%%%%%%%
\label{symm_Delta_triplet_s}
%%%%%%%%%%%%%%%%%%%%%%%%%%%%%%%%%%%%%%%%%%%%%%%%%%
\bm{\Delta}_{\mathbf{p}+1}=-\bm{\Delta}_{\mathbf{p}-1}\equiv\bm{\Delta}_{\mathbf{p}}\,.
\end{equation}
As for the intra-valley order parameters,
$\tilde{\bm{\Delta}}^{\xi\xi}_{\mathbf{p}S}$,
they should satisfy the equation
\begin{equation}
%%%%%%%%%%%%%%%%%%%%%%%%%%%%%%%%%%%%%%%%%%%%%%%%%%
\label{symm_tildeDelta_triplet}
%%%%%%%%%%%%%%%%%%%%%%%%%%%%%%%%%%%%%%%%%%%%%%%%%%
\tilde{\bm{\Delta}}^{\xi\xi}_{-\mathbf{p}S}
=
-\tilde{\bm{\Delta}}^{\xi\xi}_{\mathbf{p}S}\,.
\end{equation}
We see from this formula, that in the triplet state, the
$\tilde{\bm{\Delta}}^{\xi\xi}_{\mathbf{p}S}$
cannot be independent of the direction of
$\mathbf{p}$,
it changes sign when
$\mathbf{p}\to-\mathbf{p}$.
Thus, in the triplet state, the intra-valley order parameter should have
nodes on the Fermi surface sheets. This will increase the energy of the
triplet state (when
$\tilde{\bm{\Delta}}^{\xi\xi}_{\mathbf{p}S}\neq0$)
in comparison to the singlet one. We further expect that higher
condensation energy indicates lower transition temperature for a BCS-like
state.

In principle, self-consistency equations for
$\bm{\Delta}_{\mathbf{p}\xi}$
and
$\tilde{\bm{\Delta}}^{\xi\xi}_{\mathbf{p}S}$
can be derived in the most general form, without assuming specific
${\bf p}$-dependencies
for the order parameters. Yet, even for the most comprehensive
self-consistency conditions, only two types of solutions were discovered:
either
$\bm{\Delta}_{\mathbf{p}\xi}\neq0$,
$\tilde{\bm{\Delta}}_{\mathbf{p}\xi}=0$,
or
$\bm{\Delta}_{\mathbf{p}\xi}=0$,
$\tilde{\bm{\Delta}}_{\mathbf{p}\xi}\neq0$.
This pattern matches perfectly the structure of solutions for
Eqs.~(\ref{DeltaEq})
and~(\ref{tildeDeltaEq}),
which describe the singlet order parameters.

The solution
$\bm{\Delta}_{\mathbf{p}\xi}\neq0$,
$\tilde{\bm{\Delta}}_{\mathbf{p}\xi}=0$
has the same condensation energy as the similar singlet solution. This is a
result of our choice for the interaction functions
$U_{\mathbf{pp}'}^{\xi\xi'}$
in the form of
Eq.~\eqref{FS_rings},
where they are non-zero only in the rings around each Fermi-surface sheet.
Any possible interaction between different valleys (e.g., long-range
Coulomb interaction) that couples
$\Delta_{\mathbf{p}+1}$
and
$\Delta_{\mathbf{p}-1}$ (or $\bm{\Delta}_{\mathbf{p}+1}$
and
$\bm{\Delta}_{\mathbf{p}-1}$)
will lift this degeneracy.

As for the triplet intra-valley solution
$\bm{\Delta}_{\mathbf{p}\xi}=0$,
$\tilde{\bm{\Delta}}_{\mathbf{p}\xi}\neq0$,
this type of order parameter changes its sign several times when the
momentum $\mathbf{p}$ goes around
${\bf K}_\xi$.
For our model
interaction~\eqref{FS_rings},
the term proportional 
to~${\cos(\phi_{\mathbf{p}}-\phi_{\mathbf{p}'})}$
in
Eq.~(\ref{xi_xi})
makes such a superconducting phase possible. However, for our interaction
choice, the intra-valley triplet is less stable than an intra-valley
singlet state: both these states rely on the same set of
${\Tilde U}^{(\nu,\nu')}$
for their stability, but, due to the triplet order parameter having nodes
on the Fermi surface, the triplet is metastable relative to the nodeless
singlet.

\section{Discussion}
%%%%%%%%%%%%%%%%%%%%%%%%%%%%%%%%%%%%%%%%%%%%%%%%%%
\label{Discussion}
%%%%%%%%%%%%%%%%%%%%%%%%%%%%%%%%%%%%%%%%%%%%%%%%%%

In this paper, we examined the spin density wave and superconductivity as
potential ground states of AA-stacked bilayer graphene. At the heart of our
analysis is the effective Coulomb interaction, which was estimated using
the RPA. These calculations demonstrated that this type of coupling
is sufficient to stabilize the SDW phase in undoped AA-BLG. The
corresponding N{\'e}el temperature was estimated to be in the tens of
Kelvin which is consistent with the observed ordering in other
graphene-based  
materials~\cite{veligura2012,af_1princ2013,Velasco2012gap2mev,Bao2012,
twist_exp_insul2018}.

As for superconductivity, we expect it to appear at finite doping (when the
stronger SDW is destroyed), but the situation is not as clear and requires
additional assumptions. Although weak inter-layer attraction does emerge in
the RPA effective interaction, the Coulomb-only mechanism is insufficient
to induce an observable superconducting order parameter. Thus, to realize
the superconducting ground state, other sources of attraction, such as
phonons, must be incorporated into the formalism. While the screened
Coulomb interaction cannot stabilize an observable superconducting phase,
it serves as a kind of ``selection rule", strongly favoring inter-layer
superconducting order parameters, which is a consequence of the prominent
disparity between the inter-layer and intra-layer interactions.

Then, we studied superconductivity with the inter-layer order parameter
using a BCS-type approach. We showed that two different types of
superconductive coupling could be observed in AA-BLG, depending on the
electron-phonon interaction. The first type is the usual, zero-momentum
Cooper pairs. This occurs when the coupling between electrons with momenta
near different Dirac cones dominates or when trigonal warping is
sufficiently pronounced. Contrarily, in the second type, the Cooper pairs
can have a non-zero momentum, 
$2 \mathbf{K}_{\xi}$,
which occurs when the interaction between electrons near the same Dirac
cones is larger and the strength of trigonal warping can be neglected
(relative to the superconductivity energy scale). Naturally, the
superconductivity with non-zero Cooper pair momentum has observable
experimental consequences and could be a subject of further analysis.

We do not offer here an estimate for the superconducting transition
temperature
$T_{\rm sc}$.
We believe that such an estimate might be premature and potentially
unreliable. There are several reasons for this judgment. First, observe
that, in general,
$T_{\rm sc}$
depends on the interplay of the Coulomb repulsion and phonon-mediated
attraction. In its simplest version, such an interplay is expressed as
$\lambda_{\rm eff} = \lambda - \mu^*$,
where
$\lambda_{\rm eff}$
is the effective BCS coupling constant,
$\lambda > 0$
represents phonon-mediated attraction, and
$\mu^* > 0$
is  the pseudopotential describing the superconductivity suppression caused
by the Coulomb repulsion.

When we use
$\lambda_{\rm eff}$
to evaluate, for example, the critical temperature according to
$T_{\rm sc} \sim e^{-1/\lambda_{\rm eff}}$,
we observe that inaccuracies in our knowledge about the effective coupling
constant are amplified by the exponential non-linearity of this formula.
Unfortunately, little information about
$\lambda_{\rm eff}$
can be gathered. Both coupling constants contributing to
$\lambda_{\rm eff}$
are known to have significant error margins (for example, recent
paper~\cite{das_sarma2022supercond_acoustic}
evaluated $\lambda$ up to a factor of four, which is not that unusual in
many-body settings). Thus,
$\lambda_{\rm eff}$
accumulates even higher relative error.

In case of AB-BLG, one can partially alleviate this issue by extracting the
values from experiment (e.g.,
Ref.~\onlinecite{efetov2011supercond_theor}
attempted to evaluate $\lambda$ from photoemission data). Since
experimental data on AA-BLG are scarce, this is not an option for us.

An additional problem for AA-BLG research is that the system is less stable
than AB bilayer. Therefore, it likely requires additional stabilization
mechanisms, such as
intercalation~\cite{grubisic2023AA_exper},
which themselves may affect $\lambda$ and
$\mu^*$.
Specifically, an intercalated bilayer has additional phonon modes due to
intercalating atoms. In this respect, we note that the proposal of
Ref.~\onlinecite{das_sarma2022supercond_acoustic},
which posits the sufficiency of an acoustic phonon superconducting
mechanism for a broad range of graphene-based materials, may require
non-universal modifications to be applicable for AA-BLG.

We argue that, within our model, the superconducting order parameter is
sensitive to the doping level $x$. However, depending on the model details,
it may decrease or increase with changing $x$.

In our treatment we explicitly assumed that the doping is either zero
($x=0$),
or sufficiently strong
($x>x_c$).
This allows us to avoid the issue of an SDW/superconductivity coexistence
that may become a possibility if
$0 < x < x_c$.
The latter regime has been investigated in
Ref.~\onlinecite{FracMetal},
where we argued that AA-BLG can host several competing fractional metallic
states. Omitting certain subtleties, one can say that these states have a
well-defined Fermi surface. Consequently, under suitable conditions, they
can host superconducting phases. Recent
publications~\cite{trilayer_quarter2021exper, ab_frac2022exper}
reported experimental realization of the fractional metallicity in
graphene-based systems. Thus, a theoretical study of the coexistence regime
is a very timely endeavor. For magic-angle tBLG this coexistence was
addressed in
Ref.~\onlinecite{twisted_coexist2023},
for a more general context,
Ref.~\onlinecite{halfmetal_nesting2018prb}
may be consulted.

To conclude, we studied many-body instabilities of AA-BLG at various
dopings. We demonstrated that, at zero doping, the screened Coulomb
interaction is sufficient to induce the SDW phase with reasonable
transition temperatures. This SDW state is destroyed by doping and can be
replaced by a superconducting phase. Our calculations revealed that the
Coulomb interaction alone is insufficient for superconductivity
stabilization, and phonon-mediated attraction must be introduced. Strong
intra-layer repulsion contrasted against weak inter-layer interaction
favors inter-layer superconducting order parameters. The model supports
two types of superconducting order parameters: one type is uniform in
space, the other one demonstrates periodic spatial oscillations.  As for
the spin structure, both singlet and triplet superconductivity can be
stable.

\section*{Acknowledgments}
This work is supported by RSF grant No.~22-22-00464, \url{https://rscf.ru/en/project/22-22-00464/}.

\appendix

\section{Superconductivity via screened Coulomb interaction}
%%%%%%%%%%%%%%%%%%%%%%%%%%%%%%%%%%%%%%%%%%%%%%%%%%%
\label{AppendixA}
%%%%%%%%%%%%%%%%%%%%%%%%%%%%%%%%%%%%%%%%%%%%%%%%%%

In this Appendix we derive the formula for the superconducting coupling
constant and estimate the transition temperature for the case when the
superconductivity is induced only by the screened Coulomb interaction. For
this aim we rewrite first the single-particle Hamiltonian of the system in
the following form
\begin{eqnarray}
H_0'&=&H_0-\mu\sum_{\mathbf{k}i\alpha\sigma}
		d^\dag_{\mathbf{k}i\alpha\sigma}
		d^{\phantom{\dag}}_{\mathbf{k}i\alpha\sigma}
=
\sum_{{\bf k} \sigma}
	D^\dag_{{\bf k} \sigma}
	\hat{H}^{\vphantom{\dagger}}_{\bf k}
	D^{\vphantom{\dag}}_{{\bf k} \sigma},
\qquad
%\nonumber\\
%&&\sum_{\mathbf{k}\sigma}\sum_{ij\alpha\beta}
%d^\dag_{\mathbf{k}i\alpha\sigma}H_{\mathbf{k}}^{i\alpha;j\beta}d^{\phantom{\dag}}_{\mathbf{k}i\alpha\sigma} \label{H0new}\,.
\end{eqnarray}
where
\begin{equation}
%%%%%%%%%%%%%%%%%%%%%%%%%%%%%%%%%%%%%%%%%%%%%%%%%%
\label{D}
%%%%%%%%%%%%%%%%%%%%%%%%%%%%%%%%%%%%%%%%%%%%%%%%%%
D_{\mathbf{k}\sigma}=\left(d^{\phantom{\dag}}_{\mathbf{k}1A\sigma},d^{\phantom{\dag}}_{\mathbf{k}1B\sigma},
d^{\phantom{\dag}}_{\mathbf{k}2A\sigma},d^{\phantom{\dag}}_{\mathbf{k}2B\sigma}\right)^{\text{T}}
\end{equation}
%the function
%$H_{\mathbf{k}}^{i\alpha;j\beta}$
%can be written in the
%$4\times4$
%matrix form as [$(\hat{H}_{\mathbf{k}})^{i\alpha;j\beta}=H_{\mathbf{k}}^{i\alpha;j\beta}$]
and the matrix
$\hat{H}_{\bf k}$
is defined as
\begin{equation}
%%%%%%%%%%%%%%%%%%%%%%%%%%%%%%%%%%%%%%%%%%%%%%%%%% %
\label{Hiajb}
%%%%%%%%%%%%%%%%%%%%%%%%%%%%%%%%%%%%%%%%%%%%%%%%%% %
\hat{H}_{\mathbf{k}}
=
\left(\begin{array}{cccc}
	-\mu&-tf_{\mathbf{k}}&t_0&0\\
	-tf_{\mathbf{k}}^{*}&-\mu&0&t_0\\
	t_0&0&-\mu&-tf_{\mathbf{k}}\\
	0&t_0&-tf_{\mathbf{k}}^{*}&-\mu
\end{array}\right).
\end{equation}
Since
$f^{\vphantom{*}}_{- {\bf k}} = f^*_{\bf k}$,
matrix
$\hat{H}_{\bf k}$
satisfies the following useful relation
$\hat{H}^{\vphantom{*}}_{-\bf k} = \hat{H}^*_{\bf k}$.

Further, we rewrite the interaction Hamiltonian, keeping there only terms
relevant to the superconductivity with inter-valley order parameter
\begin{equation}
%%%%%%%%%%%%%%%%%%%%%%%%%%%%%%%%%%%%%%%%%%%%%%%%%%
\label{Hintnew}
%%%%%%%%%%%%%%%%%%%%%%%%%%%%%%%%%%%%%%%%%%%%%%%%%%
H_{\text{int}}
\!=\!
\frac{1}{2\cal N}\!\sum_{\mathbf{kk}'\sigma\sigma'\atop ij\alpha\beta}
	d^{\dag}_{\mathbf{k}i\alpha\sigma}
	d^{\dag}_{-\mathbf{k}j\beta\sigma'}
	V^{ij}_{\mathbf{k}-\mathbf{k}'}
	d^{\phantom{\dag}}_{-\mathbf{k}'j\beta\sigma'}
	d^{\phantom{\dag}}_{\mathbf{k}'i\alpha\sigma}\,,
\end{equation}
where $V^{ij}_{\mathbf{p}}$ is the effective Coulomb interaction, see Eqs.~\eqref{V11} and~\eqref{V12}.

A possible superconducting state can be either of singlet or triplet type.
Let us not specify first the type of ordering. In arbitrary superconducting
state some of the following anomalous averages are non-zero
\begin{equation}
%%%%%%%%%%%%%%%%%%%%%%%%%%%%%%%%%%%%%%%%%%%%%%%%%%
\label{app::eta}
%%%%%%%%%%%%%%%%%%%%%%%%%%%%%%%%%%%%%%%%%%%%%%%%%%
\eta_{\mathbf{k}\sigma\sigma'}^{i\alpha;j\beta}
=
\left\langle
	d^{\phantom{\dag}}_{\mathbf{k}i\alpha\sigma}
	d^{\phantom{\dag}}_{-\mathbf{k}j\beta\sigma'}
\right\rangle\,.
\end{equation}
These anomalous averages are defined in the basis of $d^{\phantom{\dag}}_{\mathbf{k}i\alpha\sigma}$ operators, instead of $\gamma^{\phantom{\dag}}_{\mathbf{p}S\xi\sigma}$ operators, Eq.~\eqref{supercond_eta_gen}. Here, we choose this representation in order to directly separate the inter-layer and intra-layer Cooper pairs. We also introduce the order parameters in the form
\begin{equation}
%%%%%%%%%%%%%%%%%%%%%%%%%%%%%%%%%%%%%%%%%%%%%%%%%%
\label{Deltaiajb}
%%%%%%%%%%%%%%%%%%%%%%%%%%%%%%%%%%%%%%%%%%%%%%%%%%
\Delta_{\mathbf{k}\sigma\sigma'}^{i\alpha;j\beta}=-\frac{1}{\cal N}\sum_{\mathbf{k}'}V^{ij}_{\mathbf{k}-\mathbf{k}'}
\eta_{\mathbf{k}'\sigma\sigma'}^{i\alpha;j\beta}\,.
\end{equation}
Using these quantities one can write the mean-field version of the interaction Hamiltonian~\eqref{Hintnew} in the form
\begin{eqnarray}
%%%%%%%%%%%%%%%%%%%%%%%%%%%%%%%%%%%%%%%%%%%%%%%%%%
\label{HintnewMF}
%%%%%%%%%%%%%%%%%%%%%%%%%%%%%%%%%%%%%%%%%%%%%%%%%%
H_{\text{int}}^{\text{MF}}
=
\frac12\sum_{\mathbf{k}\sigma\sigma'}\sum_{ij\alpha\beta}
	\left(
		d^{\dag}_{\mathbf{k}i\alpha\sigma}
		d^{\dag}_{-\mathbf{k}j\beta\sigma'}
		\Delta_{\mathbf{k}\sigma\sigma'}^{i\alpha;j\beta}
		+{\rm H.c.}
	\right)-
\nonumber
\\
\frac{1}{2{\cal N}}\sum_{\mathbf{kk}'\sigma\sigma'}\sum_{ij\alpha\beta}
\eta_{\mathbf{k}\sigma\sigma'}^{i\alpha;j\beta*}V^{ij}_{\mathbf{k}-\mathbf{k}'}\eta_{\mathbf{k}'\sigma\sigma'}^{i\alpha;j\beta}\,.
\end{eqnarray}
Since
$\hat{H}^{\vphantom{*}}_{-\bf k} = \hat{H}^*_{\bf k}$,
we can rewrite the total mean-field Hamiltonian as
\begin{equation}
%%%%%%%%%%%%%%%%%%%%%%%%%%%%%%%%%%%%%%%%%%%%%%%%%%
\label{HMFtot}
%%%%%%%%%%%%%%%%%%%%%%%%%%%%%%%%%%%%%%%%%%%%%%%%%%
H^{\text{MF}}=\frac12\sum_{\mathbf{k}}{\cal D}^{\dag}_{\mathbf{k}}\hat{\cal H}_{\mathbf{k}}{\cal D}^{\phantom{\dag}}_{\mathbf{k}}
+\frac12\!\sum_{\mathbf{k}\sigma\sigma'\atop}\sum_{ij\alpha\beta}
\eta_{\mathbf{k}\sigma\sigma'}^{i\alpha;j\beta*}\Delta_{\mathbf{k}\sigma\sigma'}^{i\alpha;j\beta}-4\mu{\cal N}\,.
\end{equation}
The
$16\times16$
matrix
$\hat{\cal H}_{\mathbf{k}}$
in this equation can be expressed in the block-matrix form as
\begin{equation}
%%%%%%%%%%%%%%%%%%%%%%%%%%%%%%%%%%%%%%%%%%%%%%%%%%
\label{HbDG}
%%%%%%%%%%%%%%%%%%%%%%%%%%%%%%%%%%%%%%%%%%%%%%%%%%
\hat{\cal H}_{\mathbf{k}}=\left(\begin{array}{cccc}
\hat{H}_{\mathbf{k}}&0&\hat{\Delta}_{\mathbf{k}\uparrow\uparrow}&\hat{\Delta}_{\mathbf{k}\uparrow\downarrow}\\
0&\hat{H}_{\mathbf{k}}&\hat{\Delta}_{\mathbf{k}\downarrow\uparrow}&\hat{\Delta}_{\mathbf{k}\downarrow\downarrow}\\
\hat{\Delta}^{\dag}_{\mathbf{k}\uparrow\uparrow}&\hat{\Delta}^{\dag}_{\mathbf{k}\downarrow\uparrow}&-\hat{H}_{\mathbf{k}}&0\\
\hat{\Delta}^{\dag}_{\mathbf{k}\uparrow\downarrow}&\hat{\Delta}^{\dag}_{\mathbf{k}\downarrow\downarrow}&0&-\hat{H}_{\mathbf{k}}
\end{array}\right),
\end{equation}
where
$(\hat{\Delta}_{\mathbf{k}\sigma\sigma'})^{i\alpha;j\beta}=\Delta_{\mathbf{k}\sigma\sigma'}^{i\alpha;j\beta}$,
and
\begin{equation}
{\cal D}_{\mathbf{k}}=\left(D^{\phantom{\dag}}_{\mathbf{k}\uparrow},D^{\phantom{\dag}}_{\mathbf{k}\downarrow},
D^{\dag}_{-\mathbf{k}\uparrow},D^{\dag}_{-\mathbf{k}\downarrow}\right)^{\text{T}}
\end{equation}
is a 16-component operator-valued vector.

The structure of the order parameter
$\Delta_{\mathbf{k}\sigma\sigma'}^{i\alpha;j\beta}$
depends on the type of the order parameter under discussion. Let us
consider first a singlet pairing. In this case one can write
$\Delta_{\mathbf{k}\sigma\sigma'}^{i\alpha;j\beta}
=
\Delta_{\mathbf{k}}^{i\alpha;j\beta}(i\sigma_y)_{\sigma\sigma'}$,
where the functions
$\Delta_{\mathbf{k}}^{i\alpha;j\beta}$
satisfy the condition
$\Delta_{-\mathbf{k}}^{i\alpha;j\beta}=\Delta_{\mathbf{k}}^{j\beta;i\alpha}$.
The structure of the function
$\Delta_{\mathbf{k}}^{i\alpha;j\beta}$
should guarantee (i)~opening of the gap at the Fermi level, and (ii)~the
stability of the corresponding superconducting phase.

Since intra-layer Coulomb interaction is always repulsive, the intra-layer
components
$\Delta_{\mathbf{k}}^{i\alpha;i\beta}$
are small. Below we will assign
$\Delta_{\mathbf{k}}^{i\alpha;i\beta}=0$.
As for inter-layer order parameter, in principle, all
$\Delta_{\mathbf{k}}^{i\alpha;j\beta}$
with
$i\neq j$
can be non-zero.

The goal of this Appendix is just to estimate the transition temperature
controlled solely by the renormalized Coulomb interaction. In order to
avoid excessive complications, let us choose the following ansatz for the
inter-layer order parameter
$\Delta_{\mathbf{k}}^{1\alpha;2\beta}
=
\Delta_{\mathbf{k}}^{2\alpha;1\beta}
=
\delta_{\alpha\beta}\Delta_{\mathbf{k}}$.
With this assumption, the order parameter
$\Delta_{\mathbf{k}}^{i\alpha;j\beta}$
can be written as a matrix
\begin{equation}
%%%%%%%%%%%%%%%%%%%%%%%%%%%%%%%%%%%%%%%%%%%%%%%%%%%
\label{DeltaSCiajb}
%%%%%%%%%%%%%%%%%%%%%%%%%%%%%%%%%%%%%%%%%%%%%%%%%%%
\hat{\Delta}_{\mathbf{k}}=\left(\begin{array}{cccc}
0&0&\Delta_{\mathbf{k}}&0\\
0&0&0&\Delta_{\mathbf{k}}\\
\Delta_{\mathbf{k}}&0&0&0\\
0&\Delta_{\mathbf{k}}&0&0
\end{array}\right).
\end{equation}
Diagonalization of the
Hamiltonian~\eqref{HbDG}
with the order
parameter~\eqref{DeltaSCiajb}
gives the quasiparticle spectrum
\begin{equation}
%%%%%%%%%%%%%%%%%%%%%%%%%%%%%%%%%%%%%%%%%%%%%%%%%%
\label{SpecSC}
%%%%%%%%%%%%%%%%%%%%%%%%%%%%%%%%%%%%%%%%%%%%%%%%%%
E^{(S)}_{\mathbf{k}}=\mp\sqrt{(t_{\mathbf{k}}\mp t_0\mp\mu)^2+\Delta_{\mathbf{k}}^2}\,,
\end{equation}
where
$t_{\mathbf{k}}=t|f_{\mathbf{k}}|$
and
$S=1,\,2,\dots,\,8$.
The spectrum is degenerate on the electron spin.

As our next simplification, let us consider only the case, when the
chemical potential is close to the transition from SDW to superconducting
state,
$\mu\sim\mu_c=\Delta^{\text{SDW}}_{\xi0}/2$.
Since
$\mu_c\ll t_0$,
we can take
$\mu=0$
in
Eq.~\eqref{SpecSC},
neglecting the difference in the Fermi momenta of two bands crossing the
Fermi level.

Minimizing the grand potential corresponding to
Hamiltonian~\eqref{HMFtot}
with respect to
$\eta_{\mathbf{k}\sigma\sigma'}^{i\alpha;j\beta}$
and taking into account the structure for the order parameter in the form
of
Eq.~\eqref{DeltaSCiajb},
we obtain the self-consistency equation for
$\Delta_{\mathbf{k}}$
at finite temperature in the form
\begin{eqnarray}
%%%%%%%%%%%%%%%%%%%%%%%%%%%%%%%%%%%%%%%%%%%%%%%%%%%
\label{GapEq}
%%%%%%%%%%%%%%%%%%%%%%%%%%%%%%%%%%%%%%%%%%%%%%%%%%%
\Delta_{\mathbf{k}}
&=&
-\frac14\sum_{\tau=\pm1}
	\int\frac{d^2\mathbf{k}'}{v_{\text{BZ}}}
            \frac{V^{12}_{\mathbf{k}-\mathbf{k}'}\Delta_{\mathbf{k}'}}
	    {\sqrt{(t_{\mathbf{k}'}-\tau t_0)^2+\Delta_{\mathbf{k}'}^2}}
\times\nonumber\\
&&\tanh\left(
	\frac{\sqrt{(t_{\mathbf{k}'}-\tau t_0)^2+\Delta_{\mathbf{k}'}^2}}
		{2T}
\right).
\end{eqnarray}
Using this equation we estimate the transition temperature by order of
magnitude. To proceed we observe that, due to the structure of
Eq.~\eqref{GapEq},
the maximum values of
$|\Delta_{\mathbf{k}}|$
are located near each Dirac point. Thus, we can introduce two
valley-specific order parameters
$\Delta_{\mathbf{p}\xi}=\Delta_{\mathbf{K}_{\xi}+\mathbf{p}}$,
which are non-zero only for
$|\mathbf{p}|<K_0$.
In the singlet state, these order parameters satisfy
Eq.~\eqref{symm_Delta_singlet}.
From
Eq.~\eqref{GapEq}
one can write approximate equation for $\Delta_{\mathbf{p}\xi}$:
\begin{eqnarray}
%%%%%%%%%%%%%%%%%%%%%%%%%%%%%%%%%%%%%%%%%%%%%%%%%%%
\label{GapEqXi}
%%%%%%%%%%%%%%%%%%%%%%%%%%%%%%%%%%%%%%%%%%%%%%%%%%%
\Delta_{\mathbf{p}\xi}&\approx &-\frac14\!\!\!\!\int\limits_{|\mathbf{p}'|<K_0}\!\!\!\!\frac{d^2\mathbf{p}'}{v_{\text{BZ}}}
\frac{V^{12}_{\mathbf{p}-\mathbf{p}'}\Delta_{\mathbf{p}'\xi}}{\sqrt{(v_{\rm F}p'-t_0)^2+\Delta_{\mathbf{p}'\xi}^2}}\times\nonumber\\
&&\tanh\left(\frac{\sqrt{(v_{\rm F}p'-t_0)^2+\Delta_{\mathbf{p}'\xi}^2}}{2T}\right).
\end{eqnarray}
Proceeding from
Eq.~\eqref{GapEq}
to
Eq.~\eqref{GapEqXi}
we (i)~omitted the inter-valley coupling restricting the integration over
$\mathbf{p}'$,
(ii)~neglect the contribution from the bands not crossing the Fermi level
(terms with $\tau=-1$), and (iii)~replace the spectrum
$t_{\mathbf{k}}$
by the linear function
$t_{\mathbf{K}_{\xi}+\mathbf{p}}\approx v_{\rm F}p$.
Simplifying even further, we replace the function
$V^{12}_{\mathbf{p}-\mathbf{p}'}$
in the integral by its value at the Fermi surface. Since at small transfer
momentum
$\mathbf{q}$
the function
$V^{12}_{\mathbf{q}}$
depends only on the absolute value of
$\mathbf{q}$,
we can introduce the function
$V^{12}(q)=V^{12}_{\mathbf{q}}$.
Assuming a specific angular dependence of the order parameter
$\Delta_{\mathbf{p}\xi}$
the integral over the radial component of
$\mathbf{p}$
can be evaluated.

Consider first the $s$-wave pairing such that
$\Delta_{\mathbf{p}\xi}=\Delta_{\xi}$
is constant for
$|\mathbf{p}|<K_0$,
and $\Delta_{\bar{\xi}}=\Delta_{\xi}$.
Performing the integration over $p$, we obtain the familiar estimate for
the transition temperature
$T_{\rm sc}\sim \varepsilon_{\rm F} e^{-1/\lambda_s}$
that is valid as long as the coupling constant
\begin{equation}
%%%%%%%%%%%%%%%%%%%%%%%%%%%%%%%%%%%%%%%%%%%%%%%%%%
\label{TcS}
%%%%%%%%%%%%%%%%%%%%%%%%%%%%%%%%%%%%%%%%%%%%%%%%%%
\lambda_s
=
-\frac{t_0}{4\pi^2\sqrt{3}t^2}\!\!
\int\limits_{0}^{2\pi}\!\!
	d\varphi\,V^{12}\!
	\left(2k_{\rm F}\left|\sin\frac{\varphi}{2}\right|\right)
\end{equation}
remains positive. Using our results for the screened Coulomb interaction,
for
$\epsilon=1$
we obtain
$\lambda_s\approx -0.016<0$.
Thus, even though the inter-layer interaction is seemingly attractive
[see
Fig.~\ref{FigV}\,(b)]
in some transferred-momentum range, it is not enough to stabilize such a
superconducting state.

However, if we choose different structure for the orbital part of
$\Delta_{\mathbf{p}\xi}$,
finite transition temperature can be recovered. Indeed, consider singlet
inter-valley order parameter that
$\Delta_{\mathbf{p}\xi}=\Delta_{\xi}\cos(\varphi_{\mathbf{p}})$
for
$|\mathbf{p}|<K_0$
and
$\Delta_{\mathbf{p}\xi}=0$,
otherwise. The function
$\Delta_{\mathbf{p}\xi}$
has two nodes at the Fermi surface sheet. According to
Eq.~\eqref{symm_Delta_singlet},
now we have
$\Delta_{\bar{\xi}}=-\Delta_{\xi}$.

As we will show below, the sign of the inter-layer interaction (i.e.,
repulsion or attraction) is not that important for this choice of the order
parameter. More important is the growth of
$V^{12}(q)$
for growing $q$ in the range
$q<2k_{\rm F}$.
Performing the same manipulations as above, we recover the transition
temperature formula
$T_{\rm sc}\sim\varepsilon_{\rm F}e^{-1/\lambda_p}$,
where
\begin{equation}
%%%%%%%%%%%%%%%%%%%%%%%%%%%%%%%%%%%%%%%%%%%%%%%%%%
\label{LambdaP}
%%%%%%%%%%%%%%%%%%%%%%%%%%%%%%%%%%%%%%%%%%%%%%%%%%
\lambda_p
=
-\frac{t_0}{4\pi^2\sqrt{3}t^2}\!\!
\int\limits_{0}^{2\pi}\!\!d\varphi\,
	V^{12}\!\left(2k_{\rm F}\left|\sin\frac{\varphi}{2}\right|\right)
	\cos\varphi\,.
\end{equation}
Since
$V^{12}(q)$
increases with $q$ when
$q<2k_{\rm F}$,
the coupling constant
$\lambda_p$
is positive, and this superconducting state is stable. However, our
calculations show that
$\lambda_p\approx 0.004$
even for
$\epsilon=1$.
Thus, the transition temperature will be virtually zero. 

Let us also comment that this result for the coupling constant cannot be
meaningfully improved by changing the model parameters and doping. At
finite doping the hole and electron Fermi surfaces are different. The
formula for $\lambda_p$ in this case contains two terms originating from
different Fermi surface sheets. We found that at finite doping
$\lambda_p$
is smaller than that for $\mu=0$. For example, at
$\mu=0.3t_0$, 
we obtain
$\lambda_p\approx 0.003$.
Changing the hopping amplitude 
$t_0$ 
also cannot substantially modify these results. Indeed, by increasing
$t_0$
we increase the density of states. Yet, at the same time, the screening
grows as well. Consequently, the coupling constant $\lambda_p$ remains
approximately the same.

Smallness of
$T_{\rm sc}$
follows from the relative smallness of the density of states at the Fermi
level of the AA-BLG, which is proportional to
$t_0/t$.
Similar calculations done in
Ref.~\onlinecite{ab_supercond2023sboychakov}
for doped and biased AB-BLG showed that the screened Coulomb interaction
alone can indeed stabilize the superconductivity in this material with
experimentally observed transition temperature.

We finally discuss the triplet state. In this case the order parameter
$\Delta_{\mathbf{k}\sigma\sigma'}^{i\alpha;j\beta}$
can be written in the form
$\Delta_{\mathbf{k}\sigma\sigma'}^{i\alpha;j\beta}
=
[
	i\bm{\Delta}_{\mathbf{k}}^{i\alpha;j\beta} \bm{\sigma}\sigma_y
]_{\sigma\sigma'}$.
In principle, different components of the vector function
$\hat{\bm{\Delta}}_{\mathbf{k}}$
can have different matrix structure on indices
$i\alpha$
and
$j\beta$.
We assume here, for simplicity, that all components are described by the
single
formula~\eqref{DeltaSCiajb}.
As a result, we obtain the same equation for the order parameter as for the
singlet case,
Eq.~\eqref{GapEq},
where we should replace
$\Delta_{\mathbf{k}}\to\bm{\Delta}_{\mathbf{k}}$.

Consider now the case
$\Delta^{x}_{\mathbf{k}}=\Delta_{\mathbf{k}}$,
$\Delta^{y}_{\mathbf{k}}=\Delta^{z}_{\mathbf{k}}=0$.
In the triplet state, the function
$\Delta_{\mathbf{k}}$
should satisfy the condition
$\Delta_{-\mathbf{k}}=-\Delta_{\mathbf{k}}$.
Assuming again that
$\Delta_{\mathbf{k}}$
is non-zero only for
$\mathbf{k}$
near each Dirac point we introduce two functions
$\Delta_{\mathbf{p}\xi}$
and take the $p$-wave type ansatz for
$\Delta_{\mathbf{p}\xi}$:
$\Delta_{\mathbf{p}\xi}=\Delta_{\xi}\cos \varphi_{\mathbf{p}}$
for
$|\mathbf{p}|<K_0$ and $\Delta_{\mathbf{p}\xi}=0$,
otherwise. In contrast to the singlet state, now we have
$\Delta_{\bar{\xi}}=\Delta_{\xi}$.

Performing the same calculations we derive the estimate for the transition
temperature
$T_{\rm sc}\sim\varepsilon_{\rm F} e^{-1/\lambda_p}$
with
$\lambda_p$
given by
Eq.~\eqref{LambdaP}.
This degeneracy between singlet and triplet states can be lifted if we take
into account the coupling between order parameters in different valleys,
$\Delta_{\mathbf{p}+1}$
and
$\Delta_{\mathbf{p}-1}$.
However, the addition contribution from this coupling to the value of
$\lambda_p$
cannot drastically enhance
$T_{\rm sc}$.
Indeed, the interaction of this kind is characterized by large transferred
momentum
($\sim 2|{\bf K}_{\pm 1}|$),
implying that the corresponding coupling is weak.

\section{Intra-layer superconducting order parameter}
%%%%%%%%%%%%%%%%%%%%%%%%%%%%%%%%%%%%%%%%%%%%%%%%%%%
\label{intra_layer_appendix}
%%%%%%%%%%%%%%%%%%%%%%%%%%%%%%%%%%%%%%%%%%%%%%%%%%
Beside the inter-layer superconducting order, an intra-layer one can be
also defined. The latter emerges naturally when the intra-layer interaction
is analyzed. Indeed applying the mean-field decoupling to the intra-layer
interaction Hamiltonian we derive
\begin{eqnarray}
%%%%%%%%%%%%%%%%%%%%%%%%%%%%%%%%%%%%%%%%%%%%%%%%%%
\label{intra_MF}
%%%%%%%%%%%%%%%%%%%%%%%%%%%%%%%%%%%%%%%%%%%%%%%%%%
H_{\text{intra}}^{\text{MF}}
&=&
\frac{1}{4 {\cal N}}\sum_{\mathbf{pp}'\xi}
	V_{\mathbf{pp}'}^{11}
	\left(
		\eta^{\xi\bar\xi}_{\mathbf{p}2}
		\!+\!
		\eta^{\xi\bar\xi}_{\mathbf{p}3}
	\right)
\times
\\
\nonumber 
&&
	\left(
		\gamma^{\dag}_{-\mathbf{p}'2\bar{\xi} \downarrow}
		\gamma^{\dag}_{\mathbf{p}'2\xi\uparrow}
		\!+\!
		\gamma^{\dag}_{-\mathbf{p}'3\bar{\xi} \downarrow}
		\gamma^{\dag}_{\mathbf{p}'3\xi\uparrow}
	\right)
	+
	{\rm H.c.},
\end{eqnarray}
where we limit ourselves to singlet superconducting phase with no nodes on
the Fermi surface, with total momentum of Cooper pairs being equal to zero.
If the intra-layer order parameter is introduced
\begin{eqnarray}
%%%%%%%%%%%%%%%%%%%%%%%%%%%%%%%%%%%%%%%%%%%%%%%%%%
\label{intRA_OP}
%%%%%%%%%%%%%%%%%%%%%%%%%%%%%%%%%%%%%%%%%%%%%%%%%%
\Delta^{\rm intra}_{{\bf p} \xi}
=
\frac{1}{4{\cal N}}\sum_{\mathbf{p}'}
	V_{\mathbf{pp}'}^{11}
	\left(
		\eta^{\xi\bar\xi}_{\mathbf{p}'2}
		+
		\eta^{\xi\bar\xi}_{\mathbf{p}'3}
	\right),
\end{eqnarray}
then
Eq.~(\ref{intra_MF})
can be re-written as follows
\begin{eqnarray}
%%%%%%%%%%%%%%%%%%%%%%%%%%%%%%%%%%%%%%%%%%%%%%%%%%
\label{intra_MF_OP}
%%%%%%%%%%%%%%%%%%%%%%%%%%%%%%%%%%%%%%%%%%%%%%%%%%
H_{\text{intra}}^{\text{MF}}
\!=\!\!
\sum_{\mathbf{p}\xi}\!
	\Delta^{\rm intra}_{{\bf p} \xi}\!
	\left(\!
		\gamma^{\dag}_{-\mathbf{p}'2\bar{\xi} \downarrow}\!
		\gamma^{\dag}_{\mathbf{p}'2\xi\uparrow}
		\!+\!
		\gamma^{\dag}_{-\mathbf{p}'3\bar{\xi} \downarrow}\!
		\gamma^{\dag}_{\mathbf{p}'3\xi\uparrow}\!
	\right)
	\!+\!
	{\rm H.c.}
\quad
\end{eqnarray}
Let us now compare the intra-layer order
parameter~(\ref{intRA_OP})
with the inter-layer one. In the main text the latter is defined as
$\Delta_{{\bf p} \xi}
=
\Delta^{\xi\bar{\xi}}_{\mathbf{p}2}
-
\Delta^{\xi\bar{\xi}}_{\mathbf{p}3}$.
Thus,
\begin{eqnarray}
\Delta_{{\bf p} \xi}
=
\frac{1}{4{\cal N}}\sum_{\mathbf{p}'}
	U_{\mathbf{pp}'}^{\xi\bar\xi}
	\left(
		\eta^{\xi\bar\xi}_{\mathbf{p}'2}
		-
		\eta^{\xi\bar\xi}_{\mathbf{p}'3}
	\right).
\end{eqnarray}
We see that the most important difference between the two types of order
parameters is the relative sign between
$\eta^{\xi\bar\xi}_{\mathbf{p}2}$
and
$\eta^{\xi\bar\xi}_{\mathbf{p}3}$.

In the superconducting phase discussed in the main text the inter-layer
order parameter is finite, implying that
$\eta^{\xi\bar\xi}_{\mathbf{p}2}$
and
$\eta^{\xi\bar\xi}_{\mathbf{p}3}$
have opposite signs. Thus, the sum
$\eta^{\xi\bar\xi}_{\mathbf{p}2} + \eta^{\xi\bar\xi}_{\mathbf{p}3}$
in
Eq.~(\ref{intRA_OP})
is small (or zero), and
$\Delta^{\rm intra}_{{\bf p} \xi}$
is strongly suppressed. We conclude that for the superconducting states
discussed in the main text, the role of the intra-layer superconducting
correlations is insignificant.

On the other hand, in the intra-layer superconducting phase matrix elements
$\eta_{{\bf p} 2}^{\xi \bar\xi}$
and
$\eta_{{\bf p} 3}^{\xi \bar\xi}$
have the same sign. As a result, such a phase incurs significant positive
contribution to the condensation energy
\begin{eqnarray}
%%%%%%%%%%%%%%%%%%%%%%%%%%%%%%%%%%%%%%%%%%%%%%%%%%
\label{eq::intra_layer_contribution}
%%%%%%%%%%%%%%%%%%%%%%%%%%%%%%%%%%%%%%%%%%%%%%%%%%
\delta E
\sim
\frac{1}{4{\cal N}}\sum_{\mathbf{pp}'\xi}
	V_{\mathbf{pp}'}^{11}
	\left(
		\eta^{\xi\bar{\xi}*}_{\mathbf{p}2}
		+
		\eta^{\xi\bar{\xi}*}_{\mathbf{p}3}
	\right)
	\left(
		\eta^{\xi\bar{\xi}}_{\mathbf{p}'2}
		+
		\eta^{\xi\bar{\xi}}_{\mathbf{p}'3}
	\right),
\end{eqnarray}
due to large intra-layer repulsion. Averaging
$\eta^{\xi\bar{\xi}}_{\mathbf{p}2,3}$
over the rings around the Fermi surfaces, and neglecting certain details,
we can simplify the latter formula even further
\begin{eqnarray}
%%%%%%%%%%%%%%%%%%%%%%%%%%%%%%%%%%%%%%%%%%%%%%%%%%
\label{eq::intra_layer_contribution_approx}
%%%%%%%%%%%%%%%%%%%%%%%%%%%%%%%%%%%%%%%%%%%%%%%%%%
\delta E
\sim
{\cal N} {\bar V}^{11} \left| \eta_2 + \eta_3 \right|^2,
\end{eqnarray}
where averaged RPA repulsion is positive
${\bar V}^{11} > 0$.
We see that in the intra-layer phase both matrix elements enhance each
other, generating large positive contribution to the condensation energy.

As for the inter-layer superconductivity, as a consequence of (partial)
cancellation, the value of
$\left| \eta_2 + \eta_3 \right|$
in
Eq.~(\ref{eq::intra_layer_contribution_approx})
is small. Moreover, in the situation of zero doping there is a symmetry
between bands~2 and~3 (they have identical Fermi surfaces and identical
Fermi velocity). This guarantees that the latter cancellation is exact:
$\eta_2 + \eta_3 \equiv 0$.
Finite doping breaks this symmetry. However, in the regime of weak doping
$\mu \sim \omega_{\rm D}$,
due to smallness of
$\omega_{\rm D}$
relative to other normal-metal scales in our system, violation of this
symmetry is not strong, and one expects that deviation from the
perfect cancellation is insignificant.


\begin{thebibliography}{46}
\bibitem[{not()}]{note_on_PSLG}
\bibinfo{note}{We observe a slight deviation from this formula due to the fact
  that we consider a lattice tight-binding model instead of the continuum one.}
\end{thebibliography}
\end{document}